\documentclass[
  aps,prb,
  showpacs,
  twocolumn,
  amsmath,
  amssymb,
  floatfix,
  superscriptaddress, 
  longbibliography,
  authoryear
]{revtex4-1}

\usepackage[utf8]{inputenc}
\usepackage{graphicx}
\usepackage[dvipsnames]{xcolor}
\usepackage{sidecap}
\usepackage{booktabs}
\usepackage{colortbl}
\usepackage[T1]{fontenc}

\setcitestyle{super}

\usepackage{hyperref}

\newcommand{\reff}[1]{Fig.~\ref{fig:#1}}
\newcommand{\refq}[1]{Eq.~(\ref{eq:#1})}

\newcommand{\reft}[1]{Tab.~\ref{tab:#1}}

\newcommand{\mat}[1]{$#1$}
\newcommand{\Chi}{\protect\raisebox{2pt}{$\chi$}}
\newcommand{\vek}[1]{\mathbf{#1}}
\newcommand{\vesan}[3]{{#1}^{#2}_{#3}}
\newcommand{\expval}[1]{\left\langle #1\right\rangle}

\newcommand{\dga}{D$\Gamma$A}
\newcommand{\pade}{Pad\'{e}}
\newcommand{\victory}{{\em victory}}
\newcommand{\Uonly}{$U$-only}
\newcommand{\UVone}{$U+V_1$}

\newcommand{\real}{\text{Re}}
\newcommand{\maxi}{\text{max}}
\newcommand{\ver}{\text{ver}}
\newcommand{\bub}{\text{bub}}

\newcommand{\Nf}{N_{\hspace{-.2em}f}}

\definecolor{bblueligh}{HTML}{6BAED6}

\makeatletter
  \def\l@subsubsection#1#2{}%
\makeatother

\begin{document}

  \title{Parquet approximation for molecules: spectrum and optical conductivity of  the Pariser-Parr-Pople model}
 
  \author{P.~Pudleiner}
  \affiliation{Institute for Solid State Physics, Vienna University of Technology, 1040 Vienna, Austria}
  \affiliation{Institute of Theoretical and Computational Physics, Graz University of Technology, 8010 Graz, Austria}
  \author{P.~Thunstr\"om}
\affiliation{Department of Physics and Astronomy, Materials Theory, Uppsala University, 751 20 Uppsala, Sweden}
  \author{A.~Valli}
  \affiliation{Institute for Solid State Physics, Vienna University of Technology, 1040 Vienna, Austria}
  \author{A.~Kauch}
  \affiliation{Institute for Solid State Physics, Vienna University of Technology, 1040 Vienna, Austria}
  \author{G.~Li}
 \affiliation{School of Physical Science and Technology, ShanghaiTech University
, Shanghai 200031, China}
  \author{K.~Held}
  \affiliation{Institute for Solid State Physics, Vienna University of Technology, 1040 Vienna, Austria}
  \date{\today}
  
  \begin{abstract} 
   We study a simple model system for the conjugated $\pi$-bonds in benzene, the Pariser-Parr-Pople (PPP) model, within the parquet approximation (PA),
   exemplifying the prospects of the PA for molecules. Advantages of the PA are its polynomial scaling with the number of orbitals, and the natural calculation of one- and two-particle spectral functions as well as of response and correlation functions. 
   We find large differences in the electronic correlations in the PPP model compared to a Hubbard model with only local interactions. 
  The quasiparticle renormalization (or mass enhancement) is much weaker in the PPP than in the Hubbard model, but the static part of the self-energy enhances the band gap of the former. Furthermore, the vertex corrections to the optical conductivity are much more important in the PPP model. Because non-local interactions strongly alter the self-energy, we conclude that the PA is more suitable for calculating conjugated $\pi$-bonds in molecules than single site dynamical mean-field theory.
  \end{abstract}
  \maketitle
  
  \section{Introduction}
  \label{sec:Intro}
The calculation of strongly correlated electron systems is particularly challenging\cite{Martin16} since strong correlations imply that many-body perturbation theory (MBPT), e.g., up to first or second order in the interaction $V$, is not a reliable approximation. Even relatively small systems pose a problem. An exact diagonalization (ED)\cite{Weisse2008,Thunstrom2012,Haverkoort2001} of the Hamiltonian, also coined full configuration interaction (CI)\cite{Sherrill1999} in chemistry,  is still possible for $N={\cal O} (20)$ states (orbitals times spins). Quantum Monte Carlo simulations (QMC)\cite{Foulkes2001}, on the other hand, suffer from the Fermionic sign problem, restricting this method to  $N={\cal O} (100)$.

Against this background, the methods of choice in solid state physics,  where 
 (infinitely extended) periodic crystals are typically considered,
are
density matrix renormalization group (DMRG)\cite{White1992,Schollwock2011} and dynamical mean field theory (DMFT).\cite{Metzner1989,Georges1992a,Georges1996} The former is favorable for 
one-dimensional systems which can be typically described by  matrix product states (MPS) with relatively small matrices. The
latter becomes exact for infinite dimensions and is a good approximation for three-dimensional problems, where a  local DMFT self-energy often yields a good description. Recently, DMFT has been extended diagrammatically\cite{RMPVertex} to account for non-local correlations beyond the local DMFT self-energy; this is done in the dynamical vertex approximation (D$\Gamma$A),\cite{Toschi2007,Kusunose2006,Katanin2009} the dual fermion (DF)\cite{Rubtsov2008} and related approaches\cite{Taranto2014,Rohringer2013,Li2015,Ayral2015,Galler2016,RMPVertex}. In chemistry, on the other hand,  the coupled cluster (CC) method\cite{Coester1960,Bartlett2007} has become the  standard approach.

The CC method is a resummation of Feynman diagrams but in contrast to MBPT up to infinite orders in $V$. However, only particular diagrams  are considered. These  can be associated with single, double, or triple excitations from a (typically) Hartree-Fock background and certain copies thereof.
Taking single and double excitations into account goes under the name of CCSD, whereas also taking triple excitations into account is coined CCSDT.\cite{Bartlett2007}

Another resummation of Feynman diagrams is better known in the field of physics: the parquet approximation (PA).\cite{Bickers2004} It is based on an exact set of equations, the parquet equations (PE). The PE require as (only) input the two-particle fully irreducible vertex $\Lambda$ and the non-interacting, one-particle Hamiltonian. If we approximate $\Lambda$ by the bare interaction, $\Lambda=V$, we obtain the PA,\cite{Bickers2004} which includes not only all diagrams up to third order in $V$ (i.e., corrections are $\sim V^4$) but in addition some diagrams up to infinite order in $V$ (including all ladder diagrams).
 The parquet equations have been developed in the 60s \cite{DeDominicis1962,DeDominicis1964} and revisited in the 90s,\cite{Bickers1991} at a time when sufficiently powerful computers were not yet available. 
More recently we have seen a revival  in the context of diagrammatic extensions of DMFT, specifically in the context of D$\Gamma$A, and for disordered problems.\cite{Janis2001,Yang2009,Tam2013,Valli2015,Li2016,Janis2017,Janis2017b}

In the (parquet) D$\Gamma$A,\cite{Valli2015,Li2016} $\Lambda$ is approximated by a local fully irreducible vertex $\Lambda=\Lambda_{\rm loc}$ (supplemented by the bare non-local Coulomb interaction) which can be calculated numerically by solving an Anderson impurity problem with continuous-time quantum Monte Carlo simulations.\cite{Gull2011a,Gunacker15,Kaufmann2017,w2dynamics2018}
In this context it has also been recognized that the PA works quite well for weak-to-intermediate Coulomb interactions,\cite{Li2016} i.e., when we are still quite far below the Mott transition.
Hitherto these studies have been mainly focused on the Hubbard model with only a local Coulomb interaction both in one\cite{Valli2015} and two\cite{Li2016} dimensions.  	Notwithstanding, a non-local interaction can be added to the Hamiltonian of the system without changing the form of the parquet equations.

In the present paper, we  employ the PA for one of the simplest but nonetheless relevant models with local ($U$) and non-local interactions ($V$): the  Pariser-Parr-Pople (PPP) model\cite{Pople53,Pariser53a}  for a benzene ring. It consists of six orbitals  (one $p_z$ orbital on  each carbon site) or $N=6\times 2$ states with a nearest-neighbor hopping and both local and non-local Coulomb interactions. It has the advantage that an exact ED solution is still possible. For comparison, we show results for the six site Hubbard model\cite{Valli2015} (\Uonly{} model) which has the same Hamiltonian except for neglecting all non-local Coulomb interactions $V$, as well as for the intermediate case, the extended Hubbard model with only the nearest-neighbor interaction (\UVone{} model).

The paper is organized as follows:
In Section   \ref{sec:Model}, we introduce the PPP model including the values of the parameters used. In Section
  \ref{sec:Method} the methods employed, i.e., the PE and the PA, are briefly  discussed, as well as the necessary changes  to the
 \victory{}~\cite{Li2017} code solving the PE.
The results obtained are presented in Section
  \ref{sec:Results}: We start in  Section
  \ref{sec:TwoParticleVertexFunction} with the 
structure of the two-particle vertex function and  changes thereof when including non-local interaction in the PPP model. Further we discuss in Section \ref{sec:EigenvaluesOfTheBSE} the leading instabilities (largest susceptibilities) on the basis of the largest eigenvalues in the  corresponding channel.
Section  
  \ref{sec:SelfenergyAndSpectra} is devoted to the self-energy and the one-particle spectral function; and 
Section
  \ref{sec:PhysicalResponse} to the optical conductivity. Finally Section 
  \ref{sec:ConclusionAndOutlook} provides a brief summary and outlook.

  \section{ Pariser-Parr-Pople model}
  \label{sec:Model}

\begin{figure}[t]
 \includegraphics[width=\linewidth]{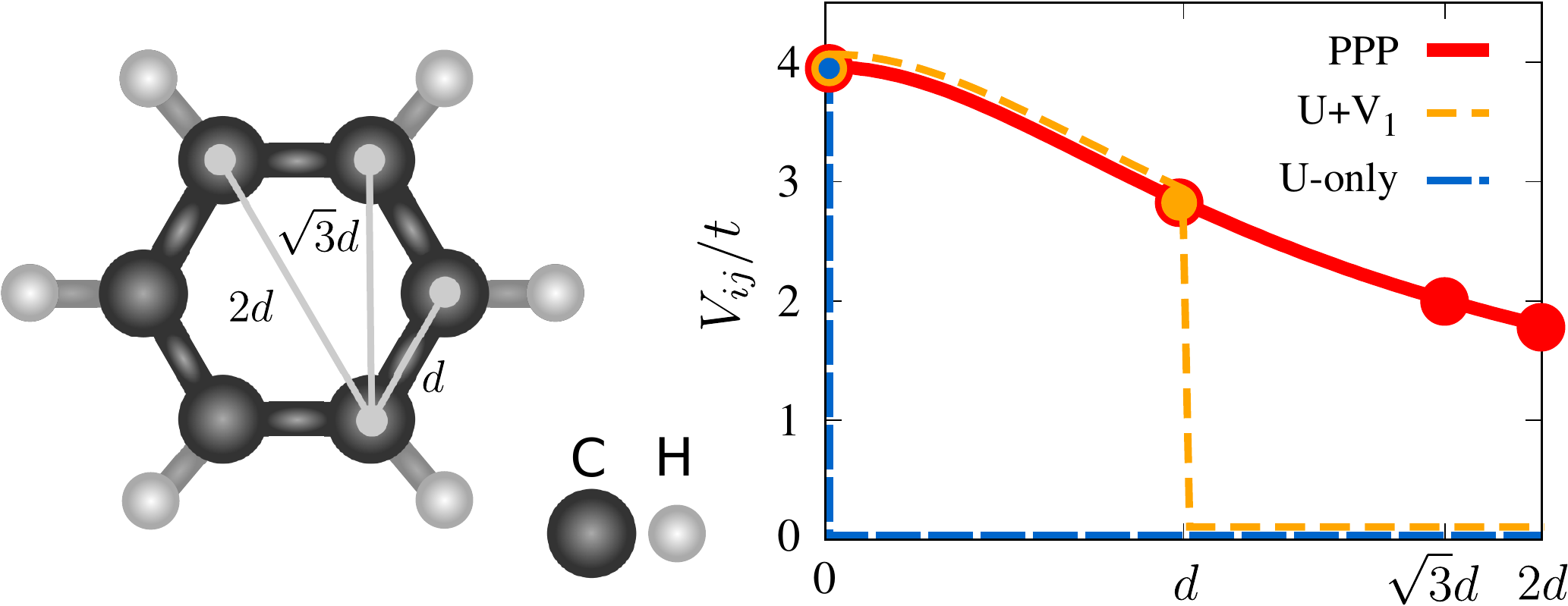}
   \caption{(Color online) Illustration of the benzene molecule, 
   chemical formula: C$_6$H$_6$ (left panel). 
   The PPP model is restricted to the $\pi$ molecular orbital 
   between the C $2p_z$ atomic orbitals. 
   The effective Coulomb interaction (right panel) is shown 
   for the different models considered: 
   (i) PPP model (red solid line), which accounts for interactions among all sites;  
   (ii) $U+V_1$ (extended Hubbard) model (dashed yellow line) 
   which accounts for an on-site and a nearest-neighbor interaction; 
    and (iii) \Uonly{} (Hubbard) model (thin blue line) 
   which restricts the interaction to its local contribution.
}
   \label{fig:benzenePPP}
\end{figure}
The Pariser-Parr-Pople (PPP) model~\cite{Pople53,Pariser53a} was developed as a realistic albeit simple one-band model of the conjugated $\pi$-bonds 
in carbon-based organic molecules. 
The model contains a $p_z$-$p_z$ hopping term $t$ between neighboring carbon sites 
and a long-range Coulomb interaction 
$V_{ij}= {U}/{\sqrt{1+\alpha |{\mathbf r}_i-{\mathbf r}_j|^2}}$. 
Here, $U$ is the local Coulomb interaction,  $\alpha=[U/(14.397\;\text{eV\AA})]^2$,
and $d_{ij}=|{\mathbf r}_i-{\mathbf r}_j|$ the distance between carbon site $i$ and $j$. 
The resulting  Hamiltonian has the form
  \begin{align}
    \mathcal{H} = - t \sum_{i\sigma}
      c^{\dagger}_{i\sigma}c_{i+1\sigma}^{\phantom{\dagger}}
      + \frac{1}{2}\! \sum_{ij, \sigma\sigma'}\! V_{ij}(n_{i\sigma}\!-\!1)(n_{j\sigma'}\!-\!1)\,,
    \label{eq:hamilton}
  \end{align}
where $n_{i\sigma}=c^{\dagger}_{i\sigma}c_{i\sigma}$, and $c^{\dagger}_{i\sigma}$ and $c_{i\sigma}$ represent creation and annihilation operators for an electron with spin $\sigma$ at carbon site  $i$, respectively.

We restrict ourselves to the PPP model for a benzene molecule (cf. left panel of \reff{benzenePPP}),
i.e., Hamiltonian (\ref{eq:hamilton}) lives on a ring with six sites with periodic boundary conditions.

The second term in \refq{hamilton} contains the effective 
density-density interaction, including the chemical potential shift
that makes the system particle-hole symmetric with $n=1$ electrons per site.
In the geometry of the benzene ring $d_{i(i+1)}=1.4$~\AA{}, $d_{i(i+2)}=\sqrt{3} \times 1.4$~\AA{}, and $d_{i(i+3)}=2 \times 1.4$~\AA{}.
Within this semiempirical approach, the parameters of the model were obtained 
in Ref.~\onlinecite{Bursill} by fits to experimental data 
of the low-lying excitation energies of benzene, yielding  
$t=2.539\,$~eV, $V_{ii} =3.962t$, $V_{i(i+1)}=2.832t$, $V_{i(i+2)}=2.014t$, and $V_{i(i+3)}=1.803t$. 
In the literature, this model for benzene has been solved 
within ED~\cite{PhysRevB.90.125413} 
and many-body perturbation theory schemes.~\cite{Tiago2005,PhysRevB.92.075422,PhysRevB.81.085102,Barford2013}

Besides the PPP model, we also study the corresponding Hubbard model (\Uonly{} model)  by setting $V_{ij}=U\delta_{ij}$, and the extended Hubbard model (\UVone{} model) with nearest-neighbor interaction with $V_{ii} =3.962t$, $V_{i(i+1)}=2.832t$ only. 
The interaction potential of these models is likewise displayed 
in the right panel of \reff{benzenePPP}.
In the following, we set our unit of energy to the hopping element, i.e., $t\equiv 1$.

The Fourier transformation to momentum (${\mathbf k}$) space yields the one-particle energies $\epsilon_{{\mathbf k}_i}=-2t \cos {\mathbf k}_i$~\cite{note_bold}, with 
${\mathbf k}_i\in\{0,\pi/3,2\pi/3,\pi,4\pi/3,5\pi/3\}$;
and interactions $V_{\vek{q}}=\sum_{l\neq0}V_{0l} e^{il\vek{q}}$. Here we have excluded the local interaction $U=V_{00}$ as it is treated separately in the following. In this work the temperature is set to $T=0.1t$.

The benzene molecule belongs to the symmetry group $D_{6h}$.\cite{doi:10.1021/ja00134a022}
The invariance of benzene under a $C_6$ rotation, 
which is an element of this group, yields a local one-particle Green's function $G_{ii}$ which is equal for all sites. 
In momentum space, the reflection symmetry of benzene 
yields the equivalence  
$\vek{k}\!=\!\frac{\pi}{3} \leftrightarrow \vek{k}=\frac{5\pi}{3}$, 
and likewise, $\vek{k}=\frac{2\pi}{3} \leftrightarrow \vek{k}=\frac{4\pi}{3}$. 
For $n\!=\!1$, the invariance under particle-hole transformation 
yields the symmetry $\vek{k}\leftrightarrow\vek{k}\pm\pi$. 
Therewith the number of independent functions reduces. 
For the one-particle functions, for instance, 
we can restrict ourselves to consider only two momenta, e.g.,  
$\vek{k}=0$ and $\vek{k}=\frac{\pi}{3}$, all other properties are obtained 
by means of the symmetries above. 

  \section{Method: extending \victory{} to treat non-local interactions}
  \label{sec:Method}

The physical quantities studied in this paper, such as transport properties or scattering rates, require the determination of both, one- and two-particle, Green's functions.
Hence, a theory that treats one- and two-particle Green's functions on an equal footing and in a self-consistent way is preferable.
To this end, we apply the parquet formalism,~\cite{Bickers_ModPhysB} which is a set of  self-consistently coupled equations, namely the Bethe-Salpeter equation, the parquet equation\cite{Diatlov1957,DeDominicis1964,DeDominicis1964b} and the Schwinger-Dyson equation.
A state-of-the-art implementation is available 
in the recent \victory{} release.~\cite{Li2017}
This package is an efficient parquet solver for electron systems in one or two dimensions, which properly takes the high-frequency asymptotics into account via so-called kernel functions.~\cite{Li2016,Wentzell2016}

While the parquet equations are an exact set of equations, they require the fully irreducible two-particle vertex $\Lambda$ as an input.
But in general $\Lambda$ is not known. 
It can be approximated by its local contribution (known as parquet-variant of the dynamical vertex approximation, \dga{}~\cite{Toschi2007,Valli2015,Li2016,RMPVertex}) which needs to be calculated by means of other methods, such as ED and QMC simulations. Or one can take the bare Coulomb interaction as $\Lambda$, 
which is also known as parquet approximation (PA).~\cite{Bickers1991,Bickers2004}

The parquet equations are numerically quite involved, 
and require an iterative solution scheme. 
In particular, since the crossing relations between vertex functions 
in different channels involve combination of momentum and frequency indices, 
the range in which the knowledge of the vertex functions is required 
increases at each iteration of the parquet equation (see below). 
Due to translational invariance (periodic boundary conditions), 
the momenta can be restricted to the first Brillouin zone. 
The frequency, however, must be kept in an, in principle, infinite range. 
The so-called kernel approximations handle this issue, 
by taking care of the vertex asymptotics in the frequency space, see Ref.~\onlinecite{Li2016}. It is beyond the scope of the present article to introduce and derive the parquet equations in full, and we refer the reader to Refs.~\onlinecite{Bickers2004,RMPVertex,Held2014,Li2016} instead.

Hitherto only a local Hubbard interaction has been treated in \victory. 
Dealing with non-local interactions requires a modification 
of the self-energy calculation by further contributions, i.e.,
the one-particle Fock term, a constant Hartree term $(U/2+V_{\mathbf q=0}) n$ 
(with the density fixed to $n=1$ electrons per site), 
and by a two-particle vertex contribution:
  \begin{eqnarray}
    \Sigma_{k}
       &=& (U/2+V_{\mathbf q=0}) n -\frac{1}{N\beta}\sum_q V_{\vek{q}}G_{q+k}
       \nonumber\\
       &&   -\frac{1}{(N\beta)^2}\sum_{k'q}
          G_{k'}G_{q+k'}G_{q+k}\nonumber\\
       &&\times \left[ \frac{U}{2} (F_{d}-F_{m})^{kk'q} + V_{\vek{q}}\vesan{F}{kk'q}{d} \right]\,.
    \label{eq:selfenergy}  \end{eqnarray}
Here, $N$ ($N=6$ for the benzene molecule) corresponds to the total number of points in the discretized momentum-space in the first Brillouin zone, and
$\beta=1/T$ is the inverse temperature.
The one-particle Green's function is labeled by $G$, and the two-particle full vertex function  by $F$.
The vertex function $F$ is directly linked to the connected part of the two-particle Green's function by cutting (``amputating'') four one-particle Green's functions from the latter.
We employ a combined four-vector notation $k=(\vek{k},\nu_n)$ for the momentum $\vek{k}$ 
and the fermionic Matsubara frequency $\nu_n=(2n+1)\pi/\beta$, 
and $q=(\vek{q},\omega_n)$ for the (one-dimensional) momentum $\vek{q}$ and the bosonic Matsubara frequencies as $\omega_n=2n\pi/\beta$, with $n\in \mathbb{Z}$.
Due to the $SU(2)$-symmetry of the Hamiltonian, the vertex function $F$ can be written in a spin-diagonalized form, with a density ($d$) and magnetic ($m$) channel; cf.~Refs.~\onlinecite{Rohringer2012,RMPVertex}.

Beside the modification in the self-energy, the momentum-dependent first-order contribution $V_{\mathbf q}$ has to be added explicitly to the fully irreducible vertex function $\Lambda$.
This is obvious for the PA, but it is also used as an approximation in the {\em ab initio} \dga{} which only takes into account local vertex corrections besides the bare interaction in $\Lambda$.~\cite{Toschi2011,Galler2016}
Both cases are treated  in our extension of the \victory{} code, by adding the bare non-local interaction to the local $\Lambda$.
In terms of formulas, we replace the bare interaction vertex $U$ in the PA by
\begin{alignat}{2}
    U^{\vek{k}\vek{k}'\vek{q}}_{d}
       &=& U 
       &\to \phantom{-}U + 2V_{\vek{q}} - V_{\vek{k}'-\vek{k}}\nonumber\\
    U^{\vek{k}\vek{k}'\vek{q}}_{m} 
       &=& -U 
       &\to -U - V_{\vek{k}'-\vek{k}}\nonumber\\
    U^{\vek{k}\vek{k}'\vek{q}}_{s} 
       &=& 2U 
       &\to \,2U + V_{\vek{q}-\vek{k}-\vek{k}'} 
          + V_{\vek{k}'-\vek{k}}\nonumber\\
    U^{\vek{k}\vek{k}'\vek{q}}_{t} 
       &=& 0 
       &\to \phantom{-} V_{\vek{q}-\vek{k}-\vek{k}'} 
          - V_{\vek{k}'-\vek{k}}\,,
   \label{eq:bareIntVertex}
\end{alignat}
where the subscripts $s$ and $t$ refer to the singlet and triplet channels which are conveniently introduced in the spin-diagonalized notation.~\cite{Li2016}

The parquet equation read,  e.g., in the density channel   (cf.~Refs.~\onlinecite{Li2016,Li2017} for all channels and further details)
\begin{eqnarray}
F_{d}^{kk^{\prime}q}&=&\Lambda_{d}^{kk^{\prime}q} + \Phi^{kk^{\prime}q}_{d} - \frac{1}{2}\Phi_{d}^{k(k+q)(k^{\prime}-k)} - \frac{3}{2}\Phi_{m}^{k(k+q)(k^{\prime}-k)} \nonumber \\ &&
+\frac{1}{2}\Phi_{s}^{kk^{\prime}(k+k^{\prime}+q)}\label{eq:PA_F}
+\frac{3}{2}\Phi_{t}^{kk^{\prime}(k+k^{\prime}+q)} \; ,
\end{eqnarray} 
where $\Phi_r$ denotes the reducible vertex function in a given parquet channel $r=d,m,s,t$.
The evaluation of the parquet equations requires frequency and momenta outside the box for which $\Phi$ is stored~\cite{Li2016}, 
due to frequency-momenta combinations such as $k^{\prime}-k$ and $k+k^{\prime}+q$ in Eq.~(\ref{eq:PA_F}).
While the periodic boundary conditions resolve this issue for the momenta, there is no periodicity in frequency space. 
For the frequencies we hence calculate the asymptotic kernel functions by scanning the surface of the known frequency box as described in  Ref.~\onlinecite{Li2016}.
However, since the bare vertex now depends on $\vek{k}$, $\vek{k}'$, and $\vek{q}$, 
so does the high frequency asymptotics of the reducible vertex functions $\Phi$. The reducible vertex function in a given channel $\Phi_r$ ($r=d,m,s,t$) is related through the Bethe-Salpeter equation to the full vertex $F$ and the irreducible vertex in channel $r$, $\Gamma_r$.  The asymptotics of $\Phi$ can thus be obtained~\cite{RMPVertex,Rohringer2013a} 
by replacing $F$ and $\Gamma$ in the Bethe-Salpeter equation by $U$, yielding
\begin{alignat}{2}
   \Phi^{kk'q}_{d/m} 
   &= &\frac{1}{N\beta} 
   &\sum_{k_1} \Gamma^{kk_1q}_{d/m} G_{k_1}G_{q+k_1}
      F^{k_1k'q}_{d/m}
   \label{eq:redVerFctdm}\\
   &\to &\frac{1}{N\beta}
   &\sum_{k_1} U^{\vek{k}\vek{k}_1\vek{q}}_{d/m} G_{k_1}G_{q+k_1}
      U^{\vek{k}_1\vek{k}'\vek{q}}_{d/m}
   \label{eq:1stredVerFctdm}\\
   \Phi^{kk'q}_{s/t}
   &= &\mp\frac{1}{2}\frac{1}{N\beta} 
   &\sum_{k_1} \Gamma^{kk_1q}_{s/t} G_{k_1}G_{q-k_1}
      F^{k_1k'q}_{s/t}
   \label{eq:redVerFctst}\\
   &\to &\mp\frac{1}{2}\frac{1}{N\beta}
   &\sum_{k_1} U^{\vek{k}\vek{k}_1\vek{q}}_{s/t} G_{k_1}G_{q+k_1}
      U^{\vek{k}_1\vek{k}'\vek{q}}_{s/t} \; .
   \label{eq:1stredVerFctst}
\end{alignat}
We utilize this high-frequency behavior when solving the Bethe-Salpeter equations by extending the frequency sum over $\nu_1$ (as part of the  $k_1$ sum) 
outside the box for which $F$ and $\Gamma$ are stored, 
i.e., using  Eqs.~(\ref{eq:1stredVerFctdm}) and (\ref{eq:1stredVerFctst}) instead of Eqs.~(\ref{eq:redVerFctdm}) and (\ref{eq:redVerFctst}). 
These terms are actually evaluated as a difference 
and in imaginary time, see Ref.~\onlinecite{Li2017}. A similar regularization (extension to high frequencies) is employed for the Dyson-Schwinger equation~(\ref{eq:selfenergy}).

With these changes we can employ \victory{} for studying non-local interactions, keeping in mind that a larger number of frequencies $\Nf$ are needed compared the case with local interactions only, due to the slower convergence to the asymptotic behavior.

  \section{Results}
  \label{sec:Results}

\begin{figure*}[t]
 \includegraphics[width=\linewidth]{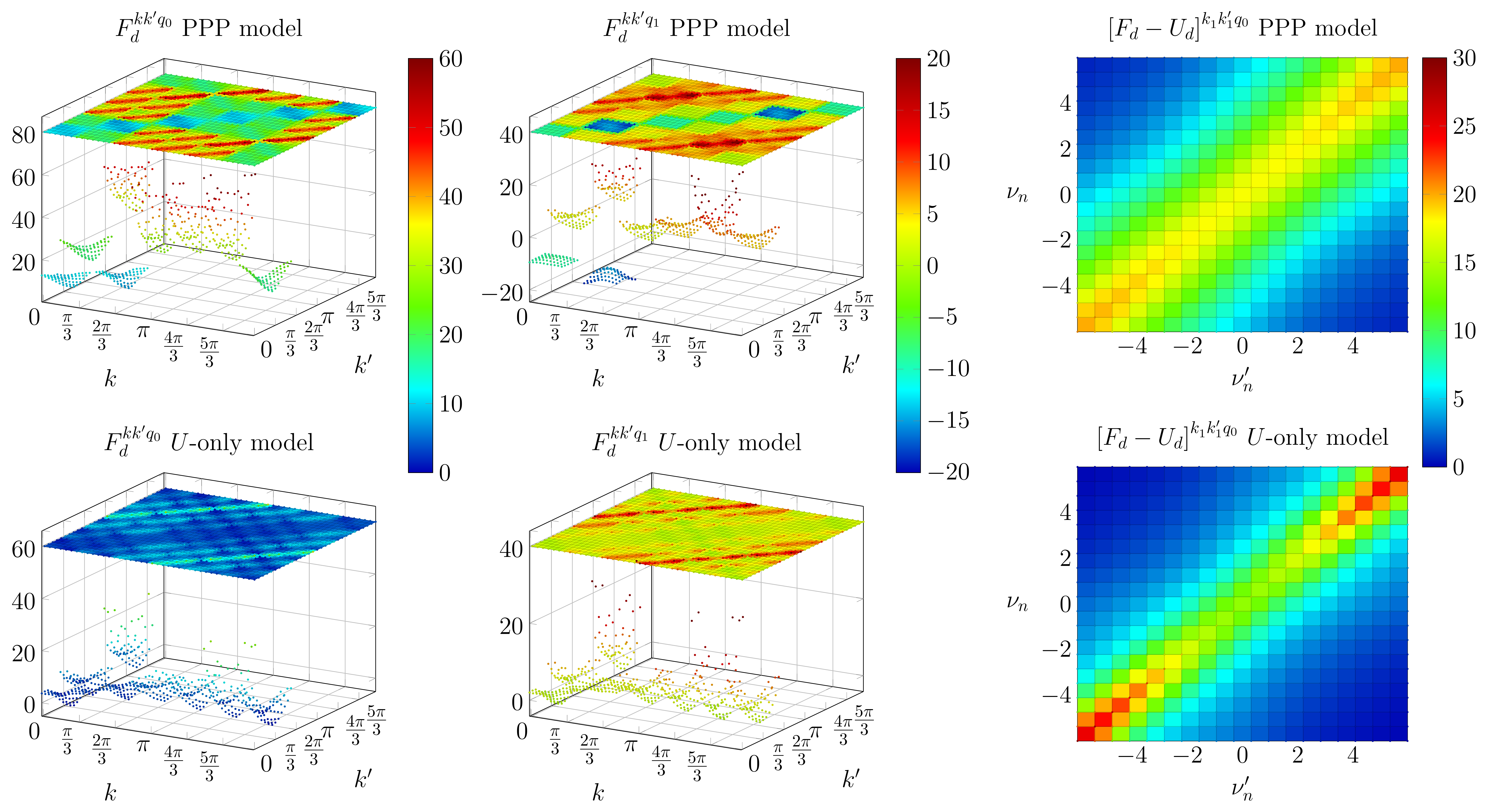}
   \caption{(Color online) Full vertex function in the density channel $F^{kk'q_0}_d$ and $F^{kk'q_1}_d$ in the PA for a benzene ring at $\beta=10/t$ and half-filling. The left and middle columns show the results  for $q_0=({\mathbf q}_0,\nu_0)=(0,0)$ and $q_1=(\pi/3,0)$, respectively, as a function of $k=({\mathbf k},\nu)$ and $k'=({\mathbf k}',\nu')$.  The top part of each panel shows
 $F^{kk'q}_d$ in a two-dimensional (2D) false-color plot with a  $\nu$, $\nu'$ subbox for each  ${\mathbf k}$, ${\mathbf k}'$.
For the following specific momentum patches a three-dimensional (3D) figure is shown within each subfigure:
 $(\vek{k},\vek{k}')= (0,\left\{0,\pi/3,\pi\right\})$ and $(\left\{0,\pi/3,2\pi/3,\pi,4\pi/3,5\pi/3\right\},\pi/3)$. 
The right column shows vertex corrections $F-U$ for $q_0$ and $k_1=(\pi,\nu)$, $k_1'=(0,\nu')$ as a function of the two Matsubara frequencies $\nu^{(\prime)}=\nu_n^{(\prime)}=(2n^{(\prime)}+1)\pi/\beta$. 
All top panels correspond to the PPP model and the bottom panels to the \Uonly{} model, as labeled.}
   \label{fig:Fd_PPP-Uonly}
\end{figure*}
%

  \subsection{Two-particle vertex function}
  \label{sec:TwoParticleVertexFunction}
Let us start by inspecting the two-particle full vertex functions which are obtained in the PA.
\reff{Fd_PPP-Uonly} shows the full vertex function $F^{kk'q}_{d}$ for the benzene ring, comparing the PPP and \Uonly{} model. 
Here, we concentrate on the density channel ($d$) since it is the relevant channel for the optical conductivity. 
The magnetic channel gives a qualitatively similar analysis.

Comparing the upper and lower row of \reff{Fd_PPP-Uonly}, we see that
including a non-local interaction in the PPP model causes a strong momentum selectivity of the full vertex function. 
The main difference between the full vertex of the PPP and \Uonly{} model stems from the additional bare non-local interaction, i.e., from $\Lambda_d = U + 2V_{\vek{q}} - V_{\vek{k}'-\vek{k}}$  in \refq{PA_F}. This frequency-independent but non-local ($\vek{q}$-dependent) interaction causes 
quite different background values for the different momentum patches (different colors in \reff{Fd_PPP-Uonly}). For the $\vek{q}=0$ contribution shown in \reff{Fd_PPP-Uonly}, this leads to a clear main diagonal structure ${\mathbf k}={\mathbf k}'$ in the vertex of the PPP model.

On top of this, in the patches with $q-k\mp k'=\textit{const.}$, we observe main and secondary diagonal structures 
also in the frequency subindex of \reff{Fd_PPP-Uonly}, 
for both models. These stem from vertex diagrams  where the external legs pairwise connect to the same interaction term,~\cite{RMPVertex} which also determine the asymptotics in the first kernel approximation, 
i.e., Eqs.~(\ref{eq:1stredVerFctdm}) and (\ref{eq:1stredVerFctst}). 
Additionally, when $q\pm k^{(\prime)}=\textit{const.}$, there is a hardly discernible plus structure which stems from vertex diagrams where only one pair of the external legs connect to the same interaction term~\cite{RMPVertex} (they constitute the second kernel function~\cite{Li2016}).

Except for the constant background given by the bare interaction, 
it seems that the frequency structure of the vertex 
for the different models is quite comparable. 
Picking for instance the momentum patch $(\vek{k},\vek{k}')=(\pi,0)$ for $q_0=(0,0)$ 
in the third column of \reff{Fd_PPP-Uonly}, 
where the constant background $U+V_{\mathbf q}$ is already subtracted, 
the frequency structures shown look similar. 
On a quantitative level, the side diagonal $\nu_n=-\nu'_n$ is somewhat sharper 
for the \Uonly{} model than for the PPP model, which instead shows a broader structure. 
The latter characteristic is the reason why an accurate simulation of the PPP model requires large frequency boxes.
This aspect is even more important for the bosonic frequency $\omega_n$, as the kernel functions are set to zero outside of the inner bosonic frequency box.

  \begin{figure}[tb]
     \centering
     \includegraphics[width=\linewidth]{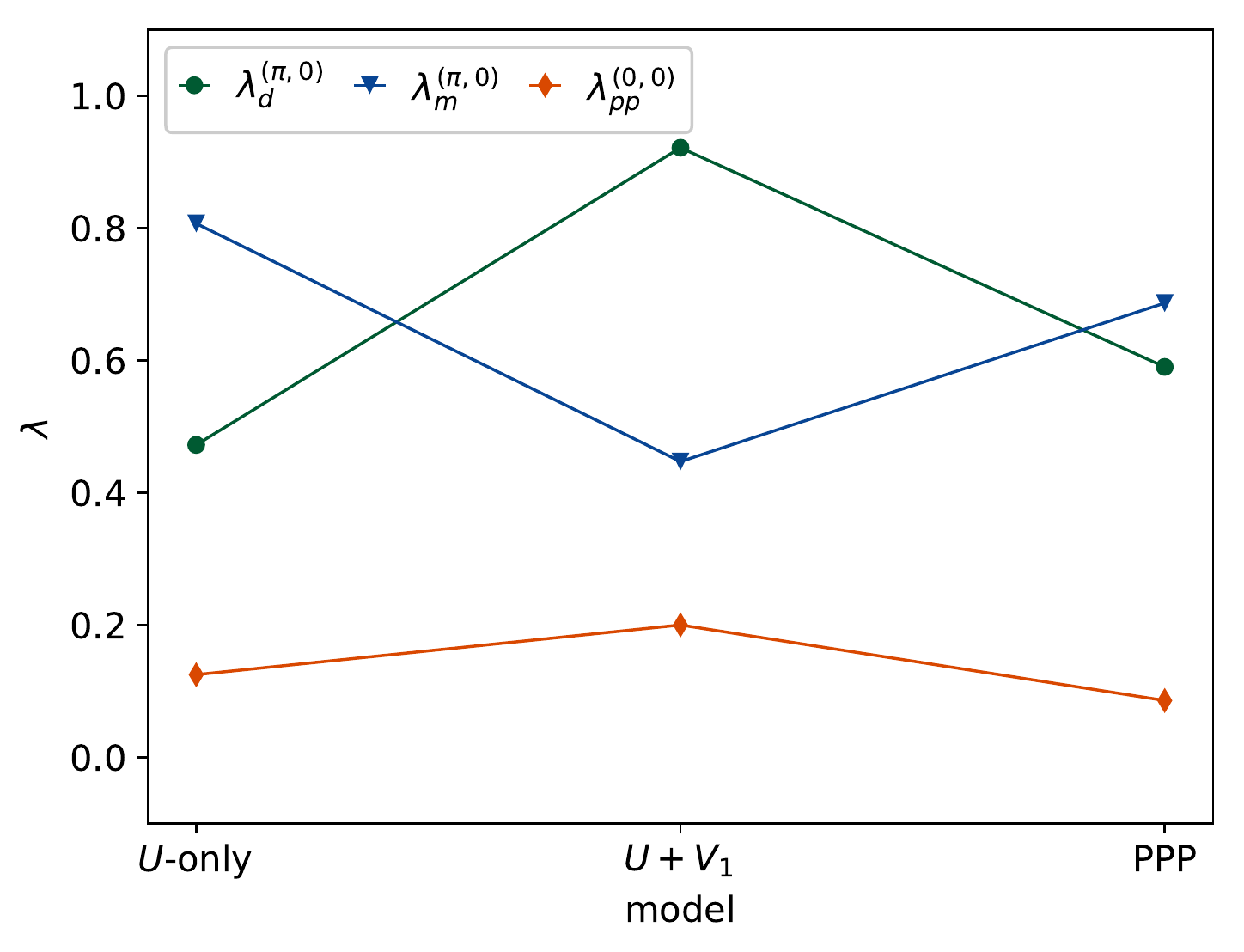}
     \caption{(Color online) Leading eigenvalues $\lambda^{q_{\maxi}}$ of \refq{EVvsV} 
     in the $d$, $m$, and $pp$ channels for the \Uonly, \UVone{} and PPP model within the PA.}
     \label{fig:EVvsV}
  \end{figure}
%
  \subsection{Eigenvalues of the BSE}
  \label{sec:EigenvaluesOfTheBSE}
Valuable insights on the physics of the benzene ring can be obtained by analyzing the eigenvalues of the Bethe-Salpeter equation, i.e., the eigenvalues $\lambda$ of 
%
\begin{eqnarray}
 \begin{aligned}
   \lambda^{q}_{d/m}\phi_{d/m}^{kq} 
      &= \frac{1}{N\beta} \sum_{k_1}\vesan{\Gamma}{kk_1q}{d/m}
         G_{k_1}G_{k_1+q} \phi_{d/m}^{k_1q} \\
   \lambda^{q}_{pp}\phi_{pp}^{kq} 
      &= \frac{1}{2N\beta} \sum_{k_1}
         \left[\Gamma_{s}-\Gamma_{t}\right]^{kk_1q}
         G_{k_1}G_{q-k_1} \phi_{pp}^{k_1q}\;.
   \label{eq:EVvsV}
 \end{aligned}
\end{eqnarray}
Here, we have for each $q$, a characteristic equation \footnote{evaluated for the $\Nf$ frequencies of the inner frequency box} for the eigenvalue $\lambda^{q}$, or more precisely three equations for the density and magnetic ($d$/$m$) channels, 
as well as for the particle-particle ($pp$) channel. For the latter the given combination of singlet and triplet ($s$/$t$) channels allows for a simultaneous study of all superconducting instabilities. 
In the following, we consider the value of $q_{\maxi}=(\vek{q}_{\maxi},\omega_{\maxi})$,   corresponding to the leading eigenvalue $\lambda^{q_{\maxi}}$ in the respective channel. 
The leading eigenvalues are shown in \reff{EVvsV}. 
For all three models considered, the dominating instabilities 
are in the $d$ and $m$ channels, which display the largest eigenvalues 
at $q_{\maxi}=(\pi,0)$, 
while the leading eigenvalues in the $pp$ channels, at $q_{\maxi}=(0,0)$, are much smaller. 

Within the \Uonly{} model, $\lambda_m$ is the largest 
among the leading eigenvalues in the different channels, 
thus revealing a strong tendency toward spin ordering, 
with an antiferromagnetic (AF) pattern associated to $\vek{q}_{\maxi}=\pi$. 
In contrast, within the \UVone{} model, the system is at the verge 
of a charge density wave (CDW) instability ($\lambda_d \approx 1$). 
The corresponding wave vector is again $\vek{q}_{\maxi}=\pi$, 
i.e., there is a strong tendency toward an alternating occupation of sites. 
Such an ordering is clearly favored by the nearest neighbor repulsive interaction $V_1$. 
The PPP model is somehow in between the previous scenarios, 
with a similarly large $\lambda$ in both the $m$ and $d$ channels, 
since the long-range tail of the Coulomb repulsion ''frustrates" both the 
AF and CDW staggered order. 
Finally, for all three models, the eigenvalue $\lambda_{pp}$ 
indicates a pairing tendency at $\vek{q}=0$. This is 
slightly enhanced for the \UVone{} model, but very far from a superconducting instability for all models.

  \begin{figure*}
     \includegraphics[width=\linewidth]{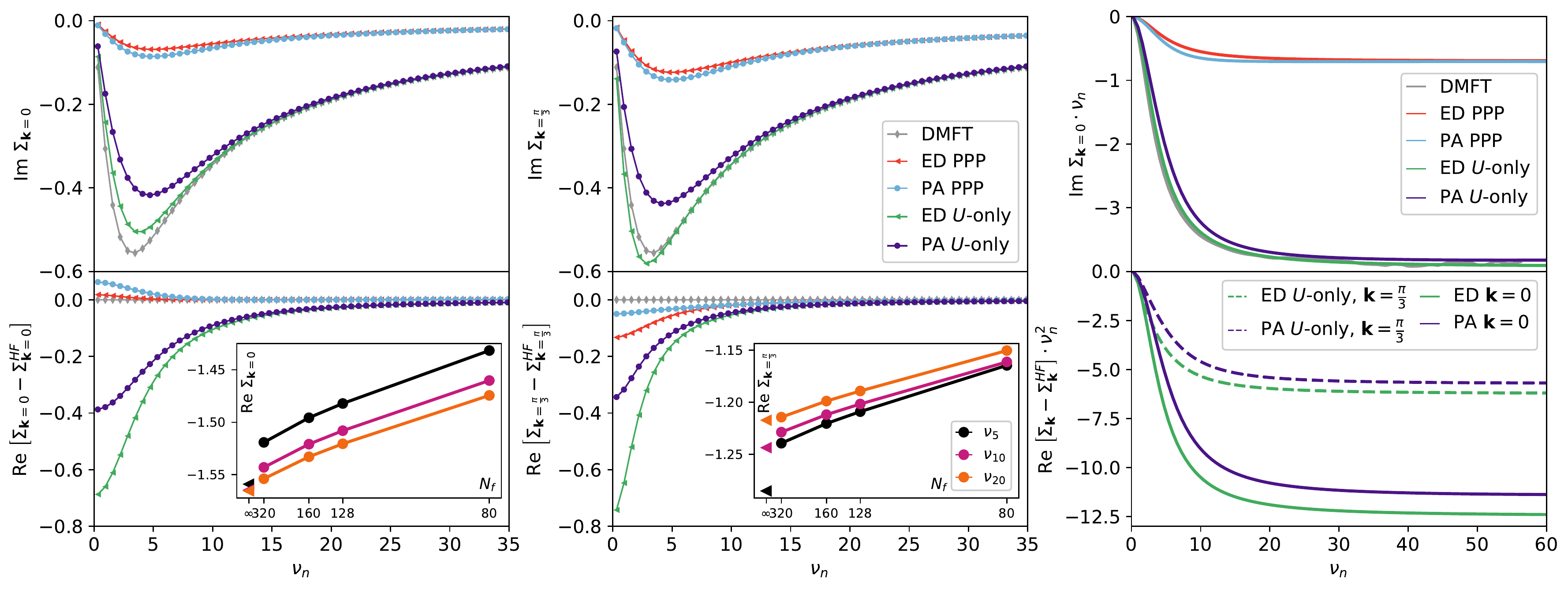}
     \caption{(Color online) Imaginary part ${\rm Im}\Sigma_{\vek{k}}$ (top panels) 
     and real part ${\rm Re}\Sigma_{\vek{k}}$ (bottom panels) of the self-energy 
     at $\vek{k}=0$ (left) and $\vek{k}=\pi/3$ (middle),  
     comparing DMFT, ED, and PA results for the \Uonly{} and PPP models. 
     In case of the PPP model, the Hartree-Fock contribution is subtracted 
     in the main figures, but the Fock term is explicitly included in the inset. 
     The inset also shows the convergence of the PA (circles) vs.~frequency box size $N_f$ 
     (note that the x-axis tics scale as $1/\Nf^2$) 
     up to $N_f=320$ (used for the data in the main panels). 
     \protect The ED results (triangles) of Ref.~\onlinecite{HoerbingerPA} are shown for comparison at  $N_f=\infty$. 
     Right panels: first (top) and second (bottom) moments 
     of the high-frequency asymptotics of the self-energy for both models 
     at $\vek{k}=0$ (top)  and the \Uonly{} model the at $\vek{k}=0$ and $\vek{k}=\pi/3$ (bottom).} 
     \label{fig:Sigmakw_PPP-Uonly}
  \end{figure*}
%
  \subsection{Self-energy and spectra}
  \label{sec:SelfenergyAndSpectra}
Let us now turn to  the self-energy shown in \reff{Sigmakw_PPP-Uonly} for the PPP model and the \Uonly{} model.
Here, the Fock (and Hartree) term is subtracted for a better comparison of the two models.
As an exact diagonalization (ED) for a one-band Hamiltonian with six sites is still feasible, the ED results for the self-energy are  displayed in \reff{Sigmakw_PPP-Uonly} as well.\cite{HoerbingerPA} 
Further we show the (${\mathbf k}$-independent) DMFT self-energy obtained by continuous time Monte Carlo simulations in the interaction expansion.\footnote{Here, we use the interaction expansion continuous time Monte Carlo (CT-QMC) method\cite{Rubtsov2005,Gull2011a} in the implementation of Gang Li.}

In the top panel of \reff{Sigmakw_PPP-Uonly}, 
we show the imaginary part of the self-energy for two $\mathbf k$-points. 
In the case of the \Uonly{} model, the DMFT self-energy provides a good description 
of the system, as it roughly corresponds to a $\mathbf k$-average of the ED data. 
The PA slightly underestimates the self-energy, but it nevertheless provides 
the correct tendency with respect to the momentum dependence, 
i.e., the self-energy is (in absolute terms) larger at $\mathbf k=(\pi/3)$ than at $\mathbf k=(0)$.  This confirms previous calculations,~\cite{Valli2013,Valli2015,Li2016} 
showing that the PA gives excellent results in the weak-to-intermediate coupling regime for the Anderson impurity model. 
Stronger deviations may instead be expected toward strong coupling. 
As long as the DMFT solution gives a reasonably correct description of the local physics, the parquet results can be improved by using the local vertex $\Lambda_{\rm loc}$ 
instead of the bare $U$ as the starting point 
for the parquet equations~\cite{Valli2013,Valli2015,Li2016}, 
i.e., using the parquet D$\Gamma$A instead of the PA. 
However, as seen in \reff{Sigmakw_PPP-Uonly}, the DMFT self-energy is clearly inaccurate for the PPP model, and a D$\Gamma$A calculation starting from a DMFT solution without self-consistency can therefore not be expected to yield reliable results.

A similar good agreement with the exact ED result 
is also obtained for the real part of the PA self-energy, 
shown in the lower panel of \reff{Sigmakw_PPP-Uonly}. 
In contrast, the real part of the self-energy is identically zero in DMFT 
because of particle-hole symmetry, 
i.e., it is qualitatively and quantitatively very different from the ED and PA result. 

When we include non-local interactions and consider the PPP model, 
the self-energy in \reff{Sigmakw_PPP-Uonly} is dramatically suppressed within ED. 
This aspect is completely missing in the DMFT, which yields exactly the same self-energy 
for both models, as long as there is no CDW order, since non-local interactions 
are only included at the Hartree level.~\cite{Muller-Hartmann1988}.
Instead, the PA provides an excellent description of the PPP model, 
and it is able to reproduce not only the overall suppression of the self-energy, 
but also the change of sign of ${\rm Re}\left[\Sigma_{\vek{k}=0}-\Sigma^{HF}_{\vek{k}=0}\right]$. 
One can rationalize the smaller self-energy for the PPP model, 
by considering the  extreme case that the non-local interaction $V_{ij}$ 
is independent of the distance $|i-j|$. 
For $V_{ij}=V_{00}=U$ and a fixed number of electrons, 
all configurations in the occupation space compatible with the Pauli exclusion principle 
have the same interaction energy. 
Hence the system behaves more like the non-interacting system. 
In particular there is neither a renormalization (narrowing) effect 
(corresponding to the large linear part of ${\rm Im}\Sigma$ for small frequencies) 
nor a different  ${\rm Re}\Sigma$ for each $\mathbf k$ points (however occupied/unoccupied states are shifted against each other). 
In the PPP model we do not yet have this extreme case, but the self-energy is already very much suppressed compared to the \Uonly{} model with the same $U$. 
As the PPP model is less strongly correlated, it is maybe not surprising that the PA provides even better results than for the  \Uonly{} model.  This suggests that the PA is a good approximation to describe the conjugated $\pi$-bonds in carbon-based molecules where non-local interactions partially compensate the effect of the local interaction. 

A separation of the self-energy in a local dynamical part and a non-local static part according to $\Sigma_{(\vek{k},\nu)}=\Sigma^{loc}_{\nu}+\Sigma^{'}_{\vek{k}}$ is discussed in Ref.~\onlinecite{Schaefer15}.
Similarly as in Ref.~\onlinecite{PhysRevBselfenergy}, we observe no clear separation of this kind.
The Fock contribution can be assigned straightforwardly to the non-local static part $\Sigma^{'}_{\vek{k}}$.
However, comparing the $\vek{k}$ and $\nu$ dependence of the self-energy in \reff{Sigmakw_PPP-Uonly}, one can notice that the separation works better for the \Uonly{} model.
Certainly, this is not the case for the PPP model, as the self-energy at different $\vek{k}$-points display a different low-frequency behavior (note the sign change of the dynamical contribution in the real part when comparing the first to the second column of \reff{Sigmakw_PPP-Uonly})

The insets of \reff{Sigmakw_PPP-Uonly} show the convergence of our results with respect to the size of the frequency box used, now including  the Fock term.
We see that at $N_f=320$ (total number of frequencies: positive and negative) deviations from the extrapolated  $N_f\rightarrow \infty$ result are very minor. Hence  $N_f=320$ has been  employed for the main panel. Also shown in the inset is the ED result\cite{HoerbingerPA} (as a triangle at  $N_f=\infty$),
and we see that PA approaches the ED results for large frequency boxes $N_f$ and high frequencies ($\nu_{20}$), whereas for smaller frequencies ($\nu_{5}$) there are deviations, as is also visible in the main panel. 
The right panel of \reff{Sigmakw_PPP-Uonly} shows that PA 
not only correctly reproduces the frequency independent Fock (and Hartree) term but also 
captures the asymptotic  $1/\nu$ behavior of the imaginary part of the self-energy (which is $\mathbf k$-independent) and the  $1/\nu^2$ behavior of the real part of the self-energy.

Reproducing the  high-frequency behavior correctly is an important non-trivial aspect of the PA and parquet variants of the D$\Gamma$A. 
In fact, within the ladder D$\Gamma$A, the correct $1/\nu$ asymptotic is only recovered if a Moriya $\lambda$-correction is included.~\cite{Katanin2009,RMPVertex} 
In dual fermion calculations,~\cite{Rubtsov2008} the correct asymptotics is obtained if the theory is truncated at the level of the two-particle vertex, but not any longer if three-particle vertex terms are included.~\cite{Katanin2013,Ribic2017b}


\reff{Aw} shows the ${\mathbf k}$-integrated and  ${\mathbf k}$-resolved 
spectral function $A_{\vek{k}}(\nu)$, which  for the PA and DMFT are obtained by an analytic continuation using the
\pade{} method \footnote{For the \pade{} interpolation, about $15$ grid points equally spaced with respect to Matsubara frequency are utilized and chosen such that the high-frequency behavior is not included.
The correct tail is anyway automatically obtained by the assumed continued fraction.}. The discrete peaks of the ED have been broadened  by a Lorentzian of width $\delta=0.05t$. Again we show results  for the PPP (right) and \Uonly{} model (left).
  \begin{figure}[tb]
     \centering
     \includegraphics[width=\linewidth]{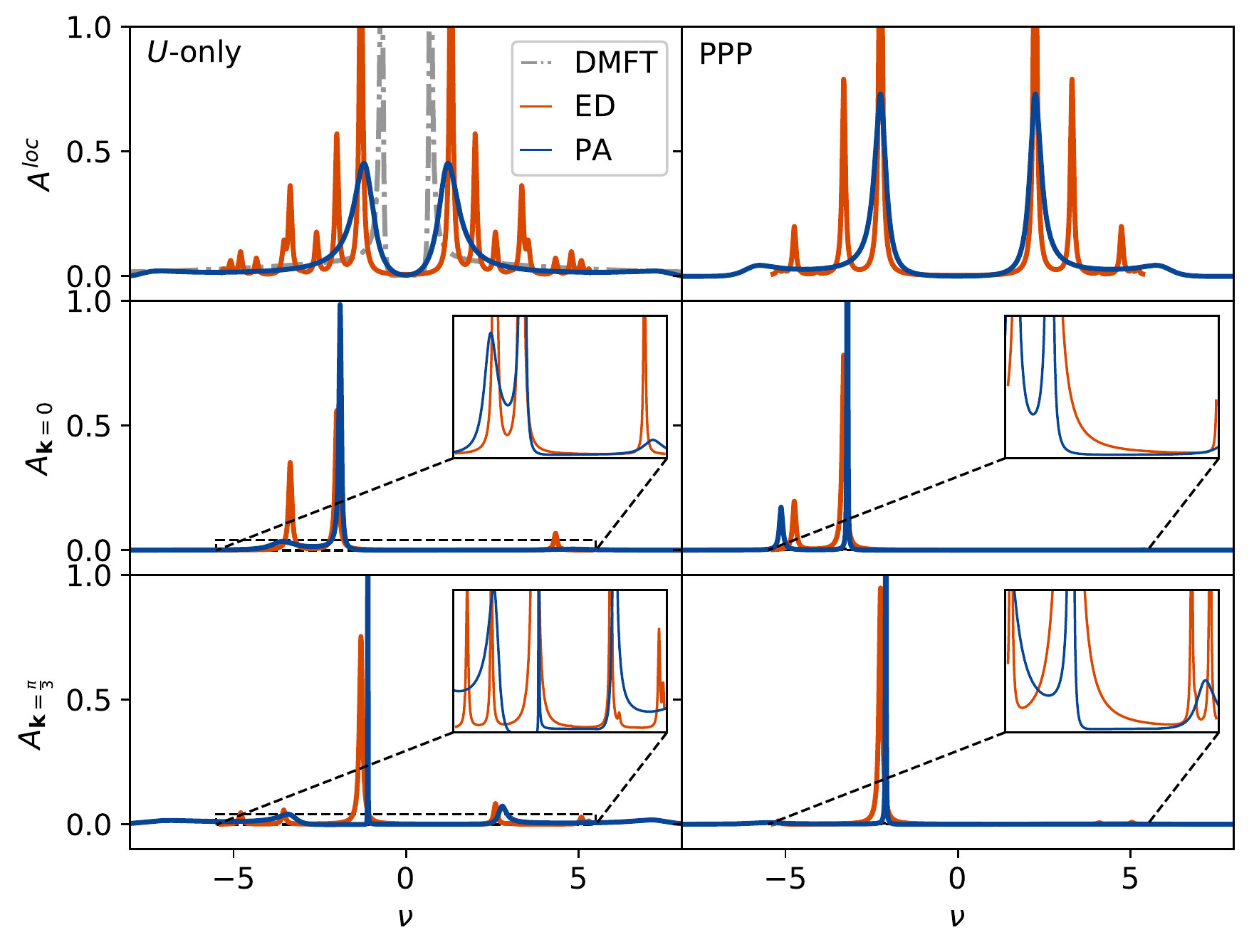}
     \caption{(Color online) Top panels: $\vek{k}$-integrated (loal) spectral function 
     $A^{\textrm{loc}}(\omega)$ obtained within DMFT, ED and PA 
     for the \Uonly{} model (left) and the PPP model (right). 
      Lower panels:  $A_{\vek{k}}(\nu)$ for $\vek{k}=0$ and $\vek{k}=\frac{\pi}{3}$,
     respectively. The inset correspond to a zoom in the indicated region.}
     \label{fig:Aw}
  \end{figure}

For the top panel of \reff{Aw}, i.e., the  ${\mathbf k}$-integrated spectral function, we see that the (HOMO-LUMO) gap is considerably larger for the PPP model than for the \Uonly{} model. To get a better understanding of this difference let us start by looking at the self-energy, which within the band gap is accurately described by its leading order terms
\begin{equation}
   \Sigma_{\vek{k}}(\nu) \approx
     \left.\real\Sigma_k\right|_{\nu\to0} + i\nu\left.\frac{\partial {\rm Im} \Sigma_k}{\partial \nu}\right|_{\nu\to0}.
\end{equation}
The static term $\real\Sigma_k(0)$, which here also includes the chemical potential, has opposite sign for the occupied part of the spectrum (${\mathbf k}=0, \pi/3$) and the unoccupied one (${\mathbf k}=2 \pi/3, \pi$). It is added to the non-interacting energies $\epsilon_{\vek{k}}$, and contributes therefore directly to the band gap. The linear term, given by the slope of ${\rm Im} \Sigma$ in \reff{Sigmakw_PPP-Uonly}, contributes instead indirectly to the band gap through the quasiparticle renormalization factor 
\begin{equation}
   Z_{\vek{k}}=
     \left[1-\left.\frac{\partial {\rm Im} \Sigma_k}{\partial \nu}\right|_{\nu\to0}\right]^{-1}\;.
\end{equation}
The two terms together yield the following positions for the poles $(\epsilon^*_{\vek{k}})$ of the one-particle Green's function
\begin{equation}
   \epsilon^*_{\vek{k}}=
     Z_{\vek{k}}\left[ \epsilon_{\vek{k}}
     +\left.\real\Sigma_k\right|_{\nu_n\to0} \right]\;.
\label{Eq:QP}
\end{equation}
The effect of $Z$ is to make the whole spectrum and hence also the band gap more narrow. Numerically, we determine the slope ${\partial {\rm Im} \Sigma_k}/{\partial \nu}$ for a finite $T$ by fitting a second order polynomial in $\nu$ to $ {\rm Im} \Sigma_{\mathbf {k},\nu_n}$.
The quasiparticle excitation energies Eq.~(\ref{Eq:QP}) nicely fit to the peak positions of the PA and ED data, as shown in the blue shaded fields of \reft{ZfactorPPP}.

Now, to understand why the gap increases for the PPP model compared to both, the  \Uonly{} model and the non-interacting spectrum, we can analyze $Z$ and $\real\Sigma_k(0)$ in \reft{ZfactorPPP}. We find $Z\approx 0.8$ for the \Uonly{} model and $Z\approx 0.95$ in the PPP model. The narrowing effect of $Z$ in the \Uonly{} model is almost completely canceled by the contribution from $\real\Sigma_k(0)$, which brings the band gap very close to its non-interacting value. In the PPP model the effect of $Z$ is instead almost negligible, and  ${\rm Re} \Sigma_k(0)$ is dominated by a large Fock term, proportional to the non-local interaction strength, which strongly enhances the band gap. Since ${\rm Re} \Sigma_k\approx 1.3$ is quite similar for ${\mathbf k}=0$ and ${\mathbf k}= \pi/3$, the effect is like cutting the non-interacting spectrum at the Fermi level with a pair of scissors and shifting the occupied (unoccupied) down (up). This scissor operator is also a major effect of the $GW$ approximation for semiconductor band gaps.\cite{Godby1988,Held2011}

\newcolumntype{g}{>{\columncolor{bblueligh}}c}
\begin{table}[tb]
\begin{ruledtabular}
\centering
\begin{tabular}{c|c|gggcc|gggcc}\toprule
\rowcolor{white}
\multicolumn{2}{c}{}&\multicolumn{5}{c}{\Uonly}&\multicolumn{5}{c}{PPP}\cr 
\hline
\rowcolor{white}
\mat{\vek{k}}&\mat{\epsilon_{\vek{k}}}  &\mat{\epsilon^*_{\vek{k}}}&\mat{\epsilon_{\text{PA}}}&\mat{\epsilon_{\text{ED}}}&\mat{\real\Sigma}&\mat{Z}&\mat{\epsilon^*_{\vek{k}}} 
  &$\epsilon_{\text{PA}}$&$\epsilon_{\text{ED}}$&\mat{\real\Sigma}&\mat{Z}\cr
\hline
\mat{0}                            &-2  &-2.0 &-1.9 &-2.0 &-0.38 & 0.83   &-3.3 &-3.1 &-3.3 &-1.47 & 0.96 \cr
\mat{\frac{\pi}{3}}                &-1  &-1.1 &-1.1 &-1.3 &-0.34 & 0.80   &-2.1 &-2.1 &-2.2 &-1.23 & 0.94 \cr
\mat{\frac{2\pi}{3}}               & 1  & 1.1 & 1.1 & 1.3 & 0.34 & 0.80   & 2.1 & 2.1 & 2.2 & 1.23 & 0.94 \cr
\mat{\pi}                          & 2  & 2.0 & 1.9 & 2.0 & 0.38 & 0.83   & 3.3 & 3.1 & 3.3 & 1.47 & 0.96 \cr
\bottomrule
\end{tabular}
\caption{Parameters of the Fermi-liquid-like renormalization, i.e., quasiparticle energy \mat{\epsilon^*_{\vek{k}}} and quasiparticle weight $Z$ compared to the non-interacting energies  \mat{\epsilon_{\vek{k}}} and the (predominant) peaks $\epsilon_{PA(ED)}$ of the PA (ED) from \reff{Aw}.
For better comparison the non-interacting dispersion relation \mat{\epsilon_{\vek{k}}} and the real part of the self-energy at $\nu_0=\pi/\beta$ is also listed.
}
\label{tab:ZfactorPPP}
\end{ruledtabular}
\end{table}

We can conclude that, despite the sometimes problematic analytic continuation procedure, 
the PA provides a very good description of the ED spectrum, 
with the spectral features being somewhat broadened. 
In particular, the PA predicts very accurately the spectral gap. 
In contrast, DMFT slightly underestimates the spectral gap for the \Uonly{} model,
which is not surprising given the previous analysis of the self-energy.

However, the DMFT results are the same for the \Uonly{} and PPP model 
(not plotted again in the right panel of \reff{Sigmakw_PPP-Uonly}), 
and thus completely miss the enhancement of the spectral gap 
due to the $\mathbf k$-dependence of  ${\rm Re} \Sigma_k$ in case of the PPP model.

The good description of the PA is even more explicit when we inspect the 
$\vek{k}$-dependent spectral function in the lower of the two plots of \reff{Aw}.
The insets therein show a zoom of the respective main panels, 
and hence resolve all the spectral features.~\footnote{Note that 
the broadening of the PA spectral function in the top panel 

is larger than for each $\vek{k}$-resolved component, 
since the \pade{} fit was performed after the $\vek{k}$-summation.} 
We observe that 
the binding energy of all the spectral features of the ED calculation are excellently reproduced 
by the $\vek{k}$-resolved components within the PA.~\footnote{The weight of the spectral features depends 
slightly on the details of the \pade{} fit procedure.}  Please note that, the spectral function at  $\vek{k}=\pi$ and $\vek{k}=\frac{2\pi}{3}$ (above the Fermi energy)
can be obtained by mirroring (at the Fermi energy the spectrum of  $\vek{k}=0$ and at $\vek{k}=\frac{\pi}{3}$.
Overall, we can record that the PA provides an excellent approximation of the one-particle spectrum. 

%
%
  \subsection{Optical conductivity}
  \label{sec:PhysicalResponse}

We study the linear response of the benzene molecule to an electric field. 
The response is related to the perturbation by the relation 
$J(\omega) = \sigma(\omega) E(\omega)$, 
where the optical conductivity $\sigma(\omega)$ corresponds to the regular part 
of the current-current correlation function
  \begin{align}
    \sigma(\omega) 
      &= e^2 \left[\frac{\Chi_{jj,\vek{q}=0}(\omega+i\delta)
                  -\Chi_{jj,\vek{q}=0}(i\delta)}{i(\omega+i\delta)} \right].
    \label{eq:conductivity}
  \end{align}
Here, the current-current correlation function $\Chi_{jj,q}$ is defined as
  \begin{align}
  \Chi_{jj,q} = \int_{0}^{\beta} d\tau 
                  \expval{j_{\vek{q}}(\tau)j_{-\vek{q}}(0)} 
                  e^{i\omega_n\tau} ,
  \end{align}
where
  \begin{align}
   j_{\vek{q}}(\tau) = & \sum_{\vek{k}} \left[ e^{-i(\vek{k}+\vek{q})}-e^{i\vek{k}} \right]c_{\vek{k}+\vek{q}}^\dagger(\tau)c_{\vek{k}}(\tau)\;.
  \end{align}
The current-current correlation function$\Chi_{jj,q}$ is evaluated in terms of the Green's function and the full vertex as
  \begin{eqnarray}
    \begin{aligned}
      \Chi_{jj,q} 
       &= \Chi^{\bub}_{jj,q} + \Chi^{\ver}_{jj,q} \\
       &= \frac{2}{\beta N}\sum_{k}
          \left[\gamma_{\vek{k}}^{\vek{q}}\right]^2
          G_{q+k}G_{k} \\
       &+ \frac{2}{(\beta N)^2}\sum_{k,k'}
          \gamma_{\vek{k}}^{\vek{q}} \gamma_{{\vek{k}}'}^{\vek{q}}
          G_{k'}G_{q+k}
          F^{kk'q}_{d}
          G_{q+k'}G_{k},
      \label{eq:ccCkw}
    \end{aligned}
  \end{eqnarray}
where we have separated the contribution from the bare bubble $\Chi^{\bub}_{jj,q}$ and that from the vertex corrections $\Chi^{\ver}_{jj,q}$. 
The matrix elements $\gamma_{\vek{k}}^{\vek{q}}$ (not to be confused with the two-particle vertices of the parquet equation) depend on the applied perturbation. 
That is, it depends on the direction in which the electric field is applied and in which direction the current response is measured.

In the following, we consider the optical response for two cases: (i) The external field is a magnetic field  applied in the direction perpendicular to the benzene ring. It  generates a magnetic flux, 
and is equivalent to a circular electric field applied along the ring. 
In this case, the coupling with the electrons at small $\vek{q}$ is approximately given by the Peierls contribution 
to the (current) vertex $\gamma_{\vek{k}}=\gamma_{\vek{k}}^{\vek{q} = 0} = 2t\sin(\vek{k})\left(\equiv\partial\epsilon_{\vek{k}}/\partial \vek{k}\right)$.
The respective correlation function is labeled by $P$ (for Peierls), namely $\Chi_P \equiv \Chi_{jj}$ and likewise the optical conductivity via $\sigma_P$.
The second case (ii) is 
the constant $\gamma_{\vek{k}}$ case, for which we use the label $C$, i.e. $\Chi_C$ and  $\sigma_C$.
This case corresponds to the dynamic compressibility or an electric field perpendicular to the ring in the limit of infinitesimally small perpendicular hopping.\footnote{
Please note that applying an electric field within the plane of the benzene ring results in a much more complicated equation than Eq.~\ref{eq:ccCkw} since it breaks  translational invariance along the ring and momentum is not conserved any longer.}

Within \victory, \refq{ccCkw} can be easily evaluated. Let us stress that, 
in order to obtain reasonable results, especially for $\Chi^{\bub}_{jj}$, 
the kernel approximations are extensively exploited. 
For obtaining the optical conductivity $\sigma(\omega)$ 
we evaluate the Matsubara current-current correlation function 
and then perform a numerical analytic continuation 
$\Chi_{jj}(\omega_n)\rightarrow\Chi_{jj}(\omega+i\delta)$ via the \pade{} method. 
The analytic continuation is a delicate procedure, and in order to obtain physical result, 
only a subset of the bosonic Matsubara frequencies has been used, as shown in the insets of \reff{ccCk0e}. 
As the  temperature considered  ($\beta=10/t$) is low enough, a sparse frequency grid can be used for the interpolation without qualitative changes in the outcome.
Note, in fact, that a larger number of input points do not result in actual poles, as they are canceled by the numerator in the continued fraction. 
\footnote{We further checked  that the  peaks calculated by  the \pade{} analytical continuation reproduce our data when transformed back to imaginary frequency where each peak gives a Lorentzian with different weight (weight of the peak in real frequencies) and width (energy of the peak).}
  \begin{figure}[tb]
     \centering
     \includegraphics[width=\linewidth]{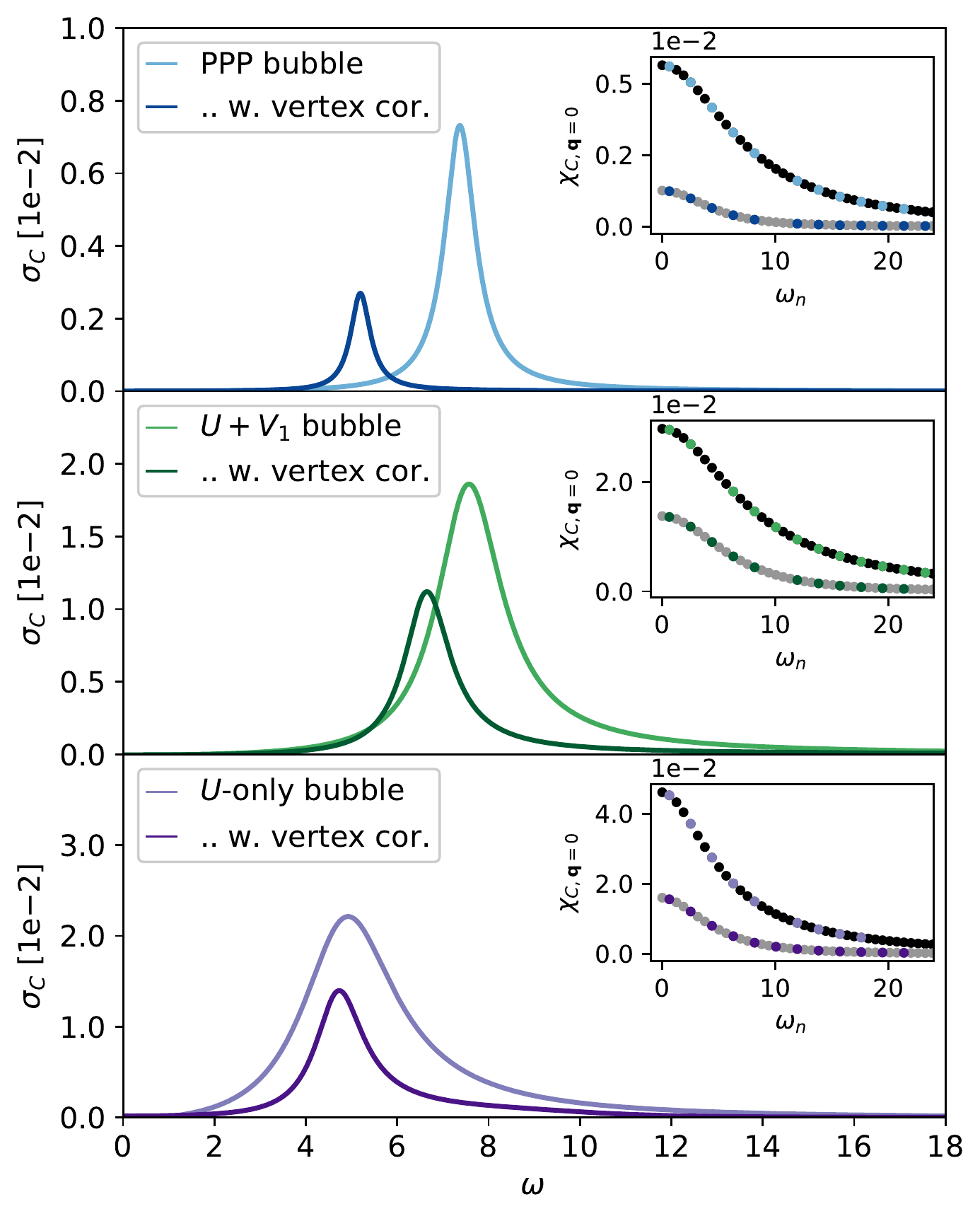}
     \caption{(Color online) Optical conductivity in $z$-direction i.e., for a constant $\gamma_{\vek{k}}$ [case (ii)],
     calculated with (dark color) and without (light color) vertex corrections 
     for the different models. 
     The insets display the corresponding current-current correlation function $\Chi_{C,\vek{q}=0}(\omega_n)$ (same color code) 
     where the colorful dots correspond to the actual grid points used for the \pade{} analytic continuation.}
     \label{fig:ccCk0e}
  \end{figure}

  
  \begin{figure}[tb]
     \includegraphics[width=\linewidth]{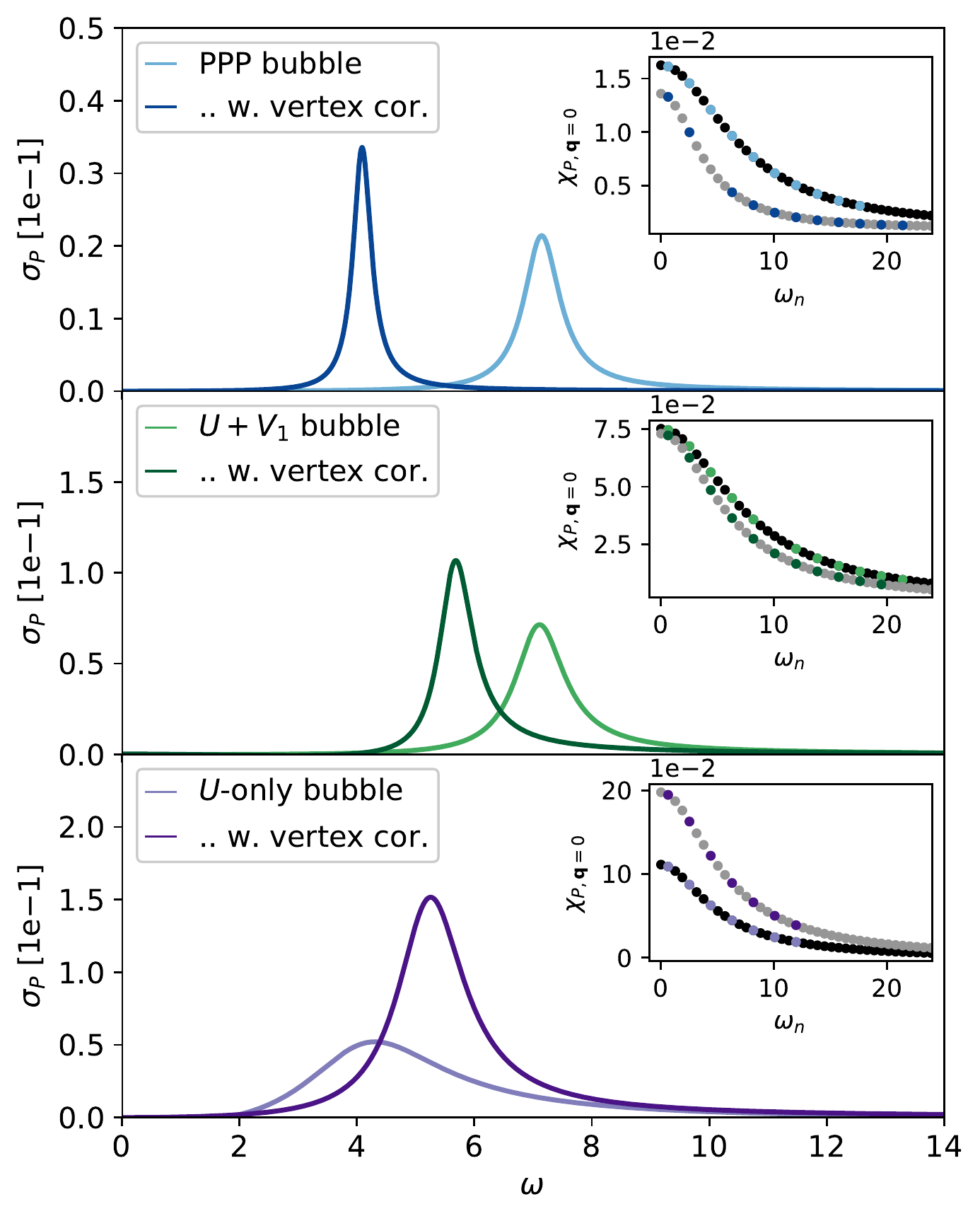}
     \caption{(Color online) Same as \reff{ccCk0e} 
     but for case (i), a perpendicular magnetic field.}
     \label{fig:opCk0e}
  \end{figure}

Let us discuss first the case (ii), i.e., the optical response to a perpendicular electric field, with $\gamma_{\vek{k}}\propto const.$ (labeled by $C$).
The results for both $\sigma_{C}(\omega)$ and $\Chi_{C,\vek{q}=0}(i\omega_n)$ 
are shown in the main panel and in the inset of \reff{ccCk0e}, respectively.
For $\Chi_{C,\vek{q}=0}(i\omega_n)$, we show the full dataset 
as well as the actual subset used for the analytic continuation. 
The optical conductivity obtained by the analytic continuation displays a single peak.
We analyze the evolution of the position of the peak at frequency $\omega^{\textrm{peak}}$ 
for the different interaction models and we observe two different effects, 
as discussed below. 
Taking into account just the bare bubble contribution $\Chi^{\bub}_C$, 
the peak if found at $\omega^{\textrm{peak}}\approx 4.9$ for the \Uonly{} model. 
It shifts to a higher frequency $\approx 7.6$ (\UVone{}) and to $\approx 7.4$ (PPP) 
when increasing the range of the Coulomb interaction. 
This is in agreement with the larger gap in the spectral function
\reff{Aw}.

The second effect is a reduction of $\omega^{\textrm{peak}}$ due to  vertex corrections. 
While this effect is relatively weak in the \Uonly{} model and still within the uncertainties of the  \pade{} fit
($\omega^{\textrm{peak}} \approx 4.8$ $\rightarrow$ $4.9$),
it becomes significantly stronger both for the \UVone{} ($7.6 \rightarrow 6.7$) 
and the PPP models ($7.4 \rightarrow 5.2$). 
This behavior is consistent with the formation of excitons 
due to non-local interactions, which cause a shift of weight in the optical conductivity 
to lower frequencies.
The fact that this effect is more pronounced 
in the PPP model than in the \UVone{} one is unexpected. 
It shows that, although the one-particle self-energy is suppressed 
by the non-local interactions in the PPP model, 
the system does not necessarily resemble an effective non-interacting limit. 
Electronic correlations induced by non-local interactions  have a much stronger effect on the two-particle response than on the one-particle properties. 

Unfortunately, a quantitative comparison of the results obtained for the PPP model 
to actual experimental data is difficult. 
The electron-impact spectra for benzene\cite{Koch1972,Lassettre1968,Doering2003} 
observe a rich structure of resonances. 
The low-energy excitations detected in the experiments in the range 
of $4.5-5.5$~eV and $6.2-6.5$~eV have been discussed to  
occur due to vibrational distortions of the benzene ring,~\cite{Doering2003} 
which are not included in our model. 
Our results for $\sigma_C(\omega)$ are instead compatible 
with the broad experimental spectrum observed at energies above $7.5$~eV. 
However, since the analytic continuation yields a single broad peak 
with a relatively large width, 
we do not have the resolution to identify any substructure.  

For the bubble contribution, we can also compare the frequency structure of $\sigma_C(\omega)$ 
to the results obtained from the one-particle Green's function 
for the local and the $\vek{k}$-resolved spectral function $A_{\vek{k}}(\omega)$  in \reff{Aw}. 
Optical transitions must occur  between occupied and unoccupied parts
of $A_{\vek{k}}(\omega)$  (i.e., between contributions below and above the Fermi level which is put to zero in \reff{Aw}). Further the optical transitions must occur between the same momentum $\vek{k}$ since the transferred momentum is
 $\vek{q}=0$.
The lowest-energy for transitions  within the \Uonly{} and PPP models 
occur both for $\vek{k}=\frac{\pi}{3}$ (and symmetrically for $\vek{k}=\frac{2\pi}{3}$), 
at a transferred energy $\Delta\omega=3.9$ and $\Delta\omega=6.9$, respectively. 
These values are off by 7\% and 20\% for the PPP model and \Uonly{} model, respectively. But given the 
ambiguity of the \pade{} method the results are actually quite comparable. 
Hence, we obtain a consistent picture between the bubble contribution to $\sigma_C(\omega)$ 
and the spectral function.

Finally, we can perform a similar analysis as the one above for the optical response 
to a perpendicular magnetic field, 
i.e., case (i) with $\gamma_{\vek{k}}=\partial\epsilon_{\vek{k}}/\partial\vek{k}$. 
The results are shown in \reff{opCk0e}. 
The peak positions of the bubble contribution to the conductivity nicely reproduce the optical gaps of the spectral function; and thus are not affected by the type of field either magnetic or electric.
Also,
the effect of vertex corrections on the position of the peak 
is qualitatively  similar to the case of perpendicular electric field, for all models. 
The main difference is that the response is overall larger [taking a dipole element $\gamma_{\mathbf k}=1$ of the same magnitude  in case (ii)]---even more so if vertex corrections are taken into account, and that the width of the main peak is substantially reduced.

%
  \section{Discussion and Conclusion}
  \label{sec:ConclusionAndOutlook}

We have studied the PPP model for the benzene molecule within the PA. 
A thorough investigation allowed us to gain insight 
of local and non-local effects on the physics of the system. 
The results obtained for the PPP model with long-range Coulomb repulsion 
are compared with those of the \UVone{} and \Uonly{} models, 
where the interaction range is restricted
up to nearest-neighbors and on-site only, respectively.  
 We solved these models within the PA, 
which account for non-local dynamical electronic correlations at all length-scales, 
as well as within the single-site dynamical mean-field theory, 
which only retains local quantum fluctuations. 
To strengthen our analysis, we further compared our results 
against the exact numerical solution, obtained by the ED of the Hamiltonian. 
Eventually, this allows us to better understand the effects 
of non-local interactions and non-local correlations within a unified formalism. 

In particular, we find that the vertex functions are 
strongly enhanced in the PPP model compared to the \Uonly{} model, 
simply because the non-local interaction $V$ itself contributes to the vertex. 
Except for this effect, both (reducible) vertices are quite similar with the vertex for the PPP model being somewhat more smeared out 
over a larger frequency range than that of the \Uonly{} model. 
As for the leading instabilities, we find that 
antiferromagnetic fluctuations are dominant for the  \Uonly{} (Hubbard) model, 
as expected. 
Whereas, charge fluctuations are so strongly enhanced for the \UVone{} model 
that this model is actually on the verge of a transition to a charge density wave. 
For the PPP model these charge fluctuations are again suppressed 
by second and third nearest neighbors repulsion, so that 
antiferromagnetic and charge density wave fluctuations turn out to be on par. 

Despite the larger vertex, we find that the imaginary part of the self-energy 
(and with this the quasiparticle renormalization) of the PPP model is actually 
much smaller than that of the \Uonly{} model. 
In this respect electronic correlations are reduced. 
This can be understood as follows. 
If we consider two electrons, they are strongly correlated in the case 
of the Hubbard (\Uonly{}) model where finding both electrons 
on the same lattice site is largely suppressed. 
The non-local interactions, on the other hand, balance the local repulsion 
and therefore mitigate differences where the two electrons are located, 
hence reducing electronic correlations. 
However, the non-local interactions also lead to a large static contribution 
to the $\mathbf k$-dependent self-energy, which in a scissors-like way 
enhances the splitting of the occupied and unoccupied states. 
For this reason the long range interactions within the PPP model 
yields a considerably larger one-particle gap
compared to the one of the Hubbard. 
For the optical conductivity we find that, for the PPP model, 
vertex corrections result in a much smaller optical gap 
than the gap in the (one-particle) spectral function. 
This is not the case for the \Uonly{} model, suggesting 
an important effect of non-local electronic correlations in the PPP model.

Since the self-energies of the PPP and \Uonly{} model are so different, DMFT is certainly not a good approximation for the PPP model (or for the conjugated $\pi$-bonds in carbon based molecules) as it yields the same self-energy (with and without non-local interactions) as long as there is no symmetry breaking. This is mitigated to some extent when using density functional theory (DFT) as a staring point in the so-called DFT+DMFT approach,\cite{Anisimov1997,Lichtenstein1998,Held2006,Kotliar2006,Held2007} but for getting the scissors-like self-energy we observe even a more involved $GW$+DMFT \cite{Biermann2003,Tomczak2012,jmt_sces14} is needed. 
This also implies that doing one-shot ({\em ab-initio}) D$\Gamma$A \cite{Galler2016,Galler2017b} calculations with the vertex taken from a DMFT impurity problem is not justified; DMFT is too far away from the correct solution. 
A self-consistent D$\Gamma$A calculation of the vertex is necessary.

Besides self-consistent D$\Gamma$A, of course also QMC, ED (CI), CC, 
and (for the special geometry considered) DMRG are viable alternatives for molecules.  Our work shows that this list should be supplemented by the PA which has a similar scaling as the CCSD, i.e., the numerical effort scales $\sim (N^2)^{2.4}$ 
($N$: number of states/orbitals; the factor of 2.4 is for an efficient matrix inversion). 
Employing symmetries and vanishing orbital off-diagonal elements, 
the effort can often be reduced, e.g., to $\sim N(N)^{2.5}$ in case of the highly symmetric molecules. 
For comparison the scaling  of CCSD  is $\sim N_u^2 N^4_o$ ($N_{u/o}$: number of unoccupied states/occupied states which are both in general $\sim N$)\cite{Bartlett2007}; and the ED even scales  exponentially with $N$. 
The biggest bottleneck of the PA is the memory 
required to store the vertex functions, which scales with 
both the number of Matsubara frequencies and states 
(or $\mathbf k$-points in case of corresponding symmetries). 

The advantages of the PA are that all one- and two-particle quantities are naturally calculated without any significant further efforts, including one-particle spectra measured experimentally in photoemission spectroscopy (PES), two-particle spectra measured e.g.,~through the optical conductivity, and 
response function as e.g.,~the magnetic susceptibility. 
For dynamical quantities (involving excited states) an analytic continuation is necessary which has some broadening effect. 
This broadening is more relevant for small molecules than for large molecules, 
where the actual spectrum is already ``smeared out'' due to the large number of energy levels. Last but not least, an advantage of the PA is that a physical understanding is fostered: we have, among others, the self-energy and vertex corrections from various channels at hand, not only the total energy and the (in general way to complicated) wave function. 

  \section*{Acknowledgment}
We would like to thank Felix H\"{o}rbinger for helping to set up the ED calculations and Tin Ribic for helpful discussions on the optical conductivity.
PP, AK and KH were supported by 
European Research Council under the 
European Union's Seventh Framework Program
(FP/2007-2013)   through ERC Grant No. 306447, and by the 
Austrian Science Fund (FWF) through SFB ViCoM F41 and project P 30997.
PP has been further supported by the FWF Doctoral School W 1243 ``Building Solids for Function''; and
AV acknowledges financial support from the Austrian Science Fund (FWF) 
through the Erwin Schr\"odinger fellowship J3890-N36, 
and through project 'LinReTraCe' P 30213. 
Calculations have been done in part on the Vienna Scientific Cluster (VSC).

  \bibliographystyle{elsarticle-num} 
  \bibliography{ref,main,addrefs2}

\begin{thebibliography}{10}
\expandafter\ifx\csname url\endcsname\relax
  \def\url#1{\texttt{#1}}\fi
\expandafter\ifx\csname urlprefix\endcsname\relax\def\urlprefix{URL }\fi
\expandafter\ifx\csname href\endcsname\relax
  \def\href#1#2{#2} \def\path#1{#1}\fi

\bibitem{Martin16}
R.~M. Martin, L.~Reining, D.~M. Ceperley, Interacting Electrons: Theory and
  Computational Approaches, Cambridge University Press Cambridge, 2016.

\bibitem{Weisse2008}
A.~{Wei{\ss}e}, H.~{Fehske}, {Exact Diagonalization Techniques}, in:
  H.~{Fehske}, R.~{Schneider}, A.~{Wei{\ss}e} (Eds.), Computational
  Many-Particle Physics, Vol. 739 of Lecture Notes in Physics, Berlin Springer
  Verlag, 2008, p. 529.
\newblock \href {http://dx.doi.org/10.1007/978-3-540-74686-7_18}
  {\path{doi:10.1007/978-3-540-74686-7_18}}.

\bibitem{Thunstrom2012}
P.~Thunstr\"om, I.~Di~Marco, O.~Eriksson,
  \href{https://link.aps.org/doi/10.1103/PhysRevLett.109.186401}{Electronic
  entanglement in late transition metal oxides}, Phys. Rev. Lett. 109 (2012)
  186401.
\newblock \href {http://dx.doi.org/10.1103/PhysRevLett.109.186401}
  {\path{doi:10.1103/PhysRevLett.109.186401}}.
\newline\urlprefix\url{https://link.aps.org/doi/10.1103/PhysRevLett.109.186401}

\bibitem{Haverkoort2001}
M.~W. Haverkort, \href{http://stacks.iop.org/1742-6596/712/i=1/a=012001}{Quanty
  for core level spectroscopy - excitons, resonances and band excitations in
  time and frequency domain}, Journal of Physics: Conference Series 712~(1)
  (2016) 012001.
\newline\urlprefix\url{http://stacks.iop.org/1742-6596/712/i=1/a=012001}

\bibitem{Sherrill1999}
C.~Sherrill, H.~F.~Schaefer, The {Configuration} {Interaction} {Method}:
  {Advances} in {Highly} {Correlated} {Approaches}, Advances in Quantum
  Chemistry 34 (1999) 143--269.
\newblock \href {http://dx.doi.org/10.1016/S0065-3276(08)60532-8}
  {\path{doi:10.1016/S0065-3276(08)60532-8}}.

\bibitem{Foulkes2001}
W.~M.~C. Foulkes, L.~Mitas, R.~J. Needs, G.~Rajagopal,
  \href{https://link.aps.org/doi/10.1103/RevModPhys.73.33}{Quantum {Monte}
  {Carlo} simulations of solids}, Rev. Mod. Phys. 73 (2001) 33--83.
\newblock \href {http://dx.doi.org/10.1103/RevModPhys.73.33}
  {\path{doi:10.1103/RevModPhys.73.33}}.
\newline\urlprefix\url{https://link.aps.org/doi/10.1103/RevModPhys.73.33}

\bibitem{White1992}
S.~R. White,
  \href{https://link.aps.org/doi/10.1103/PhysRevLett.69.2863}{Density matrix
  formulation for quantum renormalization groups}, Phys. Rev. Lett. 69 (1992)
  2863--2866.
\newblock \href {http://dx.doi.org/10.1103/PhysRevLett.69.2863}
  {\path{doi:10.1103/PhysRevLett.69.2863}}.
\newline\urlprefix\url{https://link.aps.org/doi/10.1103/PhysRevLett.69.2863}

\bibitem{Schollwock2011}
U.~Schollw\"ock, The density matrix renormalization group in the age of matrix
  product states, Ann. of Phys. 326 (2011) 96.

\bibitem{Metzner1989}
W.~Metzner, D.~Vollhardt,
  \href{http://link.aps.org/doi/10.1103/PhysRevLett.62.324}{Correlated lattice
  fermions in $d=\infty$ dimensions}, Phys. Rev. Lett. 62 (1989) 324--327.
\newblock \href {http://dx.doi.org/10.1103/PhysRevLett.62.324}
  {\path{doi:10.1103/PhysRevLett.62.324}}.
\newline\urlprefix\url{http://link.aps.org/doi/10.1103/PhysRevLett.62.324}

\bibitem{Georges1992a}
A.~Georges, G.~Kotliar,
  \href{http://link.aps.org/doi/10.1103/PhysRevB.45.6479}{Hubbard model in
  infinite dimensions}, Phys. Rev. B 45 (1992) 6479--6483.
\newblock \href {http://dx.doi.org/10.1103/PhysRevB.45.6479}
  {\path{doi:10.1103/PhysRevB.45.6479}}.
\newline\urlprefix\url{http://link.aps.org/doi/10.1103/PhysRevB.45.6479}

\bibitem{Georges1996}
A.~Georges, G.~Kotliar, W.~Krauth, M.~J. Rozenberg,
  \href{http://dx.doi.org/10.1103/RevModPhys.68.13}{Dynamical mean-field theory
  of strongly correlated fermion systems and the limit of infinite dimensions},
  Rev. Mod. Phys. 68~(1) (1996) 13.
\newblock \href {http://dx.doi.org/10.1103/RevModPhys.68.13}
  {\path{doi:10.1103/RevModPhys.68.13}}.
\newline\urlprefix\url{http://dx.doi.org/10.1103/RevModPhys.68.13}

\bibitem{RMPVertex}
G.~Rohringer, H.~Hafermann, A.~Toschi, A.~A. Katanin, A.~E. Antipov, M.~I.
  Katsnelson, A.~I. Lichtenstein, A.~N. Rubtsov, K.~Held,
  \href{https://link.aps.org/doi/10.1103/RevModPhys.90.025003}{Diagrammatic
  routes to nonlocal correlations beyond dynamical mean field theory}, Rev.
  Mod. Phys. 90 (2018) 025003.
\newblock \href {http://dx.doi.org/10.1103/RevModPhys.90.025003}
  {\path{doi:10.1103/RevModPhys.90.025003}}.
\newline\urlprefix\url{https://link.aps.org/doi/10.1103/RevModPhys.90.025003}

\bibitem{Toschi2007}
A.~Toschi, A.~A. Katanin, K.~Held, Dynamical vertex approximation; a step
  beyond dynamical mean-field theory, Phys Rev. B 75 (2007) 045118.
\newblock \href {http://dx.doi.org/10.1103/PhysRevB.75.045118}
  {\path{doi:10.1103/PhysRevB.75.045118}}.

\bibitem{Kusunose2006}
H.~Kusunose, \href{http://jpsj.ipap.jp/link?JPSJ/75/054713/}{Influence of
  spatial correlations in strongly correlated electron systems: Extension to
  dynamical mean field approximation}, J. Phys. Soc. Jpn. 75~(5) (2006) 054713.
\newblock \href {http://dx.doi.org/10.1143/JPSJ.75.054713}
  {\path{doi:10.1143/JPSJ.75.054713}}.
\newline\urlprefix\url{http://jpsj.ipap.jp/link?JPSJ/75/054713/}

\bibitem{Katanin2009}
A.~A. Katanin, A.~Toschi, K.~Held,
  \href{http://link.aps.org/doi/10.1103/PhysRevB.80.075104}{Comparing pertinent
  effects of antiferromagnetic fluctuations in the two- and three-dimensional
  {Hubbard} model}, Phys. Rev. B 80 (2009) 075104.
\newblock \href {http://dx.doi.org/10.1103/PhysRevB.80.075104}
  {\path{doi:10.1103/PhysRevB.80.075104}}.
\newline\urlprefix\url{http://link.aps.org/doi/10.1103/PhysRevB.80.075104}

\bibitem{Rubtsov2008}
A.~N. Rubtsov, M.~I. Katsnelson, A.~I. Lichtenstein, Dual fermion approach to
  nonlocal correlations in the {Hubbard} model, Phys. Rev. B 77 (2008) 033101.
\newblock \href {http://dx.doi.org/10.1103/PhysRevB.77.033101}
  {\path{doi:10.1103/PhysRevB.77.033101}}.

\bibitem{Taranto2014}
C.~Taranto, S.~Andergassen, J.~Bauer, K.~Held, A.~Katanin, W.~Metzner,
  G.~Rohringer, A.~Toschi,
  \href{http://link.aps.org/doi/10.1103/PhysRevLett.112.196402}{From infinite
  to two dimensions through the functional renormalization group}, Phys. Rev.
  Lett. 112 (2014) 196402.
\newblock \href {http://dx.doi.org/10.1103/PhysRevLett.112.196402}
  {\path{doi:10.1103/PhysRevLett.112.196402}}.
\newline\urlprefix\url{http://link.aps.org/doi/10.1103/PhysRevLett.112.196402}

\bibitem{Rohringer2013}
G.~Rohringer, A.~Toschi, H.~Hafermann, K.~Held, V.~I. Anisimov, A.~A. Katanin,
  \href{http://link.aps.org/doi/10.1103/PhysRevB.88.115112}{One-particle
  irreducible functional approach: A route to diagrammatic extensions of the
  dynamical mean-field theory}, Phys. Rev. B 88 (2013) 115112.
\newline\urlprefix\url{http://link.aps.org/doi/10.1103/PhysRevB.88.115112}

\bibitem{Li2015}
G.~Li, \href{http://link.aps.org/doi/10.1103/PhysRevB.91.165134}{Hidden physics
  in the dual-fermion approach: A special case of a nonlocal expansion scheme},
  Phys. Rev. B 91 (2015) 165134.
\newblock \href {http://dx.doi.org/10.1103/PhysRevB.91.165134}
  {\path{doi:10.1103/PhysRevB.91.165134}}.
\newline\urlprefix\url{http://link.aps.org/doi/10.1103/PhysRevB.91.165134}

\bibitem{Ayral2015}
T.~Ayral, O.~Parcollet,
  \href{http://link.aps.org/doi/10.1103/PhysRevB.92.115109}{Mott physics and
  spin fluctuations: a unified framework}, Phys Rev. B 92 (2015) 115109.
\newline\urlprefix\url{http://link.aps.org/doi/10.1103/PhysRevB.92.115109}

\bibitem{Galler2016}
A.~Galler, P.~Thunstr\"om, P.~Gunacker, J.~M. Tomczak, K.~Held,
  \href{http://link.aps.org/doi/10.1103/PhysRevB.95.115107}{Ab initio dynamical
  vertex approximation}, Phys. Rev. B 95 (2017) 115107.
\newblock \href {http://dx.doi.org/10.1103/PhysRevB.95.115107}
  {\path{doi:10.1103/PhysRevB.95.115107}}.
\newline\urlprefix\url{http://link.aps.org/doi/10.1103/PhysRevB.95.115107}

\bibitem{Coester1960}
F.~Coester, H.~Kümmel,
  \href{http://www.sciencedirect.com/science/article/pii/0029558260901401}{Short-range
  correlations in nuclear wave functions}, Nuclear Physics 17 (1960) 477 --
  485.
\newblock \href
  {http://dx.doi.org/https://doi.org/10.1016/0029-5582(60)90140-1}
  {\path{doi:https://doi.org/10.1016/0029-5582(60)90140-1}}.
\newline\urlprefix\url{http://www.sciencedirect.com/science/article/pii/0029558260901401}

\bibitem{Bartlett2007}
R.~J. Bartlett, M.~Musia\l{},
  \href{https://link.aps.org/doi/10.1103/RevModPhys.79.291}{Coupled-cluster
  theory in quantum chemistry}, Rev. Mod. Phys. 79 (2007) 291--352.
\newblock \href {http://dx.doi.org/10.1103/RevModPhys.79.291}
  {\path{doi:10.1103/RevModPhys.79.291}}.
\newline\urlprefix\url{https://link.aps.org/doi/10.1103/RevModPhys.79.291}

\bibitem{Bickers2004}
N.~E. Bickers, Theoretical {M}ethods for {S}trongly {C}orrelated {E}lectrons,
  Springer-Verlag New York Berlin Heidelberg, 2004, Ch.~6, pp. 237--296.

\bibitem{DeDominicis1962}
C.~{De Dominicis},
  \href{http://jmp.aip.org/resource/1/jmapaq/v3/i5/p983_s1}{Variational
  formulations of equilibrium statistical mechanics}, J. Math. Phys. 3~(5)
  (1962) 983--1002.
\newblock \href {http://dx.doi.org/10.1063/1.1724313}
  {\path{doi:10.1063/1.1724313}}.
\newline\urlprefix\url{http://jmp.aip.org/resource/1/jmapaq/v3/i5/p983_s1}

\bibitem{DeDominicis1964}
C.~{De Dominicis}, P.~C. Martin,
  \href{http://jmp.aip.org/resource/1/jmapaq/v5/i1/p14_s1}{Stationary entropy
  principle and renormalization in normal and superfluid systems. {I}.
  {Algebraic} formulation}, J. Math. Phys. 5~(1) (1964) 14--30.
\newblock \href {http://dx.doi.org/10.1063/1.1704062}
  {\path{doi:10.1063/1.1704062}}.
\newline\urlprefix\url{http://jmp.aip.org/resource/1/jmapaq/v5/i1/p14_s1}

\bibitem{Bickers1991}
N.~E. Bickers, S.~R. White,
  \href{http://link.aps.org/doi/10.1103/PhysRevB.43.8044}{Conserving
  approximations for strongly fluctuating electron systems. {II}. {Numerical}
  results and parquet extension}, Phys. Rev. B 43 (1991) 8044--8064.
\newblock \href {http://dx.doi.org/10.1103/PhysRevB.43.8044}
  {\path{doi:10.1103/PhysRevB.43.8044}}.
\newline\urlprefix\url{http://link.aps.org/doi/10.1103/PhysRevB.43.8044}

\bibitem{Janis2001}
V.~Jani\v{s}, \href{http://link.aps.org/doi/10.1103/PhysRevB.64.115115}{Parquet
  approach to nonlocal vertex functions and electrical conductivity of
  disordered electrons}, Phys. Rev. B 64 (2001) 115115.
\newblock \href {http://dx.doi.org/10.1103/PhysRevB.64.115115}
  {\path{doi:10.1103/PhysRevB.64.115115}}.
\newline\urlprefix\url{http://link.aps.org/doi/10.1103/PhysRevB.64.115115}

\bibitem{Yang2009}
S.~X. Yang, H.~Fotso, J.~Liu, T.~A. Maier, K.~Tomko, E.~F. D'Azevedo, R.~T.
  Scalettar, T.~Pruschke, M.~Jarrell,
  \href{http://link.aps.org/doi/10.1103/PhysRevE.80.046706}{Parquet
  approximation for the $4\ifmmode\times\else\texttimes\fi{}4$ {Hubbard}
  cluster}, Phys. Rev. E 80 (2009) 046706.
\newblock \href {http://dx.doi.org/10.1103/PhysRevE.80.046706}
  {\path{doi:10.1103/PhysRevE.80.046706}}.
\newline\urlprefix\url{http://link.aps.org/doi/10.1103/PhysRevE.80.046706}

\bibitem{Tam2013}
K.-M. Tam, H.~Fotso, S.-X. Yang, T.-W. Lee, J.~Moreno, J.~Ramanujam,
  M.~Jarrell, \href{http://link.aps.org/doi/10.1103/PhysRevE.87.013311}{Solving
  the parquet equations for the {Hubbard} model beyond weak coupling}, Phys.
  Rev. E 87 (2013) 013311.
\newblock \href {http://dx.doi.org/10.1103/PhysRevE.87.013311}
  {\path{doi:10.1103/PhysRevE.87.013311}}.
\newline\urlprefix\url{http://link.aps.org/doi/10.1103/PhysRevE.87.013311}

\bibitem{Valli2015}
A.~Valli, T.~Sch\"afer, P.~Thunstr\"om, G.~Rohringer, S.~Andergassen,
  G.~Sangiovanni, K.~Held, A.~Toschi,
  \href{http://link.aps.org/doi/10.1103/PhysRevB.91.115115}{Dynamical vertex
  approximation in its parquet implementation: Application to {Hubbard}
  nanorings}, Phys. Rev. B 91 (2015) 115115.
\newblock \href {http://dx.doi.org/10.1103/PhysRevB.91.115115}
  {\path{doi:10.1103/PhysRevB.91.115115}}.
\newline\urlprefix\url{http://link.aps.org/doi/10.1103/PhysRevB.91.115115}

\bibitem{Li2016}
G.~Li, N.~Wentzell, P.~Pudleiner, P.~Thunstr\"om, K.~Held,
  \href{https://link.aps.org/doi/10.1103/PhysRevB.93.165103}{Efficient
  implementation of the parquet equations: Role of the reducible vertex
  function and its kernel approximation}, Phys. Rev. B 93 (2016) 165103.
\newblock \href {http://dx.doi.org/10.1103/PhysRevB.93.165103}
  {\path{doi:10.1103/PhysRevB.93.165103}}.
\newline\urlprefix\url{https://link.aps.org/doi/10.1103/PhysRevB.93.165103}

\bibitem{Janis2017}
V.~Jani\v{s}, A.~Kauch, V.~Pokorn\'y, Thermodynamically consistent description
  of criticality in models of correlated electrons, Phys. Rev. B 95 (2017)
  045108.
\newblock \href {http://dx.doi.org/10.1103/PhysRevB.95.045108}
  {\path{doi:10.1103/PhysRevB.95.045108}}.

\bibitem{Janis2017b}
V.~Jani\ifmmode~\check{s}\else \v{s}\fi{}, V.~Pokorn\'y, A.~Kauch,
  \href{https://link.aps.org/doi/10.1103/PhysRevB.95.165113}{Mean-field
  approximation for thermodynamic and spectral functions of correlated
  electrons: Strong coupling and arbitrary band filling}, Phys. Rev. B 95
  (2017) 165113.
\newblock \href {http://dx.doi.org/10.1103/PhysRevB.95.165113}
  {\path{doi:10.1103/PhysRevB.95.165113}}.
\newline\urlprefix\url{https://link.aps.org/doi/10.1103/PhysRevB.95.165113}

\bibitem{Gull2011a}
E.~Gull, A.~J. Millis, A.~I. Lichtenstein, A.~N. Rubtsov, M.~Troyer, P.~Werner,
  \href{http://link.aps.org/doi/10.1103/RevModPhys.83.349}{Continuous-time
  {Monte} {Carlo} methods for quantum impurity models}, Rev. Mod. Phys. 83~(2)
  (2011) 349.
\newblock \href {http://dx.doi.org/10.1103/RevModPhys.83.349}
  {\path{doi:10.1103/RevModPhys.83.349}}.
\newline\urlprefix\url{http://link.aps.org/doi/10.1103/RevModPhys.83.349}

\bibitem{Gunacker15}
P.~Gunacker, M.~Wallerberger, E.~Gull, A.~Hausoel, G.~Sangiovanni, K.~Held,
  \href{http://link.aps.org/doi/10.1103/PhysRevB.92.155102}{Continuous-time
  quantum {Monte} {Carlo} using worm sampling}, Phys. Rev. B 92 (2015) 155102.
\newblock \href {http://dx.doi.org/10.1103/PhysRevB.92.155102}
  {\path{doi:10.1103/PhysRevB.92.155102}}.
\newline\urlprefix\url{http://link.aps.org/doi/10.1103/PhysRevB.92.155102}

\bibitem{Kaufmann2017}
J.~Kaufmann, P.~Gunacker, K.~Held,
  \href{https://link.aps.org/doi/10.1103/PhysRevB.96.035114}{Continuous-time
  quantum {Monte} {Carlo} calculation of multiorbital vertex asymptotics},
  Phys. Rev. B 96 (2017) 035114.
\newblock \href {http://dx.doi.org/10.1103/PhysRevB.96.035114}
  {\path{doi:10.1103/PhysRevB.96.035114}}.
\newline\urlprefix\url{https://link.aps.org/doi/10.1103/PhysRevB.96.035114}

\bibitem{w2dynamics2018}
M.~Wallerberger, A.~Hausoel, P.~Gunacker, F.~Goth, N.~Parragh, K.~Held,
  P.~Sangiovanni, {w2dynamics: Local one- and two-parrticle quantities from
  dynamcial mean field theory}, ArXiv e-prints\href
  {http://arxiv.org/abs/arXiv:1801.10209} {\path{arXiv:arXiv:1801.10209}}.

\bibitem{Pople53}
J.~A. Pople, \href{http://dx.doi.org/10.1039/TF9534901375}{Electron interaction
  in unsaturated hydrocarbons}, Trans. Faraday Soc. 49 (1953) 1375--1385.
\newblock \href {http://dx.doi.org/10.1039/TF9534901375}
  {\path{doi:10.1039/TF9534901375}}.
\newline\urlprefix\url{http://dx.doi.org/10.1039/TF9534901375}

\bibitem{Pariser53a}
R.~{Pariser}, R.~G. {Parr}, A {Semi-Empirical} {Theory} of the {Electronic}
  {Spectra} and {Electronic} {Structure} of {Complex} {Unsaturated}
  {Molecules}. {I}., Journal of Chemical Physics 21 (1953) 466--471.
\newblock \href {http://dx.doi.org/10.1063/1.1698929}
  {\path{doi:10.1063/1.1698929}}.

\bibitem{Li2017}
G.~{Li}, A.~{Kauch}, P.~{Pudleiner}, K.~{Held},
  \href{https://arxiv.org/abs/1708.07457}{The \emph{victory} project v1.0: an
  efficient parquet equations solver}, arXiv:1708.07457\href
  {http://arxiv.org/abs/1708.07457} {\path{arXiv:1708.07457}}.
\newline\urlprefix\url{https://arxiv.org/abs/1708.07457}

\bibitem{Bursill}
R.~J. Bursill, C.~Castleton, W.~Barford, Optimal parametrisation of the
  {Pariser-Parr-Pople} {Model} for benzene and biphenyl, Chem. Phys. Lett. 294
  (1998) 305--313.

\bibitem{PhysRevB.90.125413}
K.~G.~L. Pedersen, M.~Strange, M.~Leijnse, P.~Hedeg\aa{}rd, G.~C. Solomon,
  J.~Paaske, Quantum interference in off-resonant transport through single
  molecules, Phys. Rev. B 90 (2014) 125413.
\newblock \href {http://dx.doi.org/10.1103/PhysRevB.90.125413}
  {\path{doi:10.1103/PhysRevB.90.125413}}.

\bibitem{Tiago2005}
M.~L.~Tiago, J.~Chelikowsky, First-principles {GW-BSE} excitations in organic
  molecules, Solid State Communications 136 (2005) 333--337.
\newblock \href {http://dx.doi.org/10.1016/j.ssc.2005.08.012}
  {\path{doi:10.1016/j.ssc.2005.08.012}}.

\bibitem{PhysRevB.92.075422}
M.~P. Ljungberg, P.~Koval, F.~Ferrari, D.~Foerster, D.~S\'anchez-Portal,
  Cubic-scaling iterative solution of the {Bethe-Salpeter} equation for finite
  systems, Phys. Rev. B 92 (2015) 075422.
\newblock \href {http://dx.doi.org/10.1103/PhysRevB.92.075422}
  {\path{doi:10.1103/PhysRevB.92.075422}}.

\bibitem{PhysRevB.81.085102}
K.~Kaasbjerg, K.~S. Thygesen, Benchmarking gw against exact diagonalization for
  semiempirical models, Phys. Rev. B 81 (2010) 085102.
\newblock \href {http://dx.doi.org/10.1103/PhysRevB.81.085102}
  {\path{doi:10.1103/PhysRevB.81.085102}}.

\bibitem{Barford2013}
W.~Barford, Electronic and Optical Properties of Conjugated Polymers,
  International Series of Monographs on Physics, Oxford University Press, 2013.
\newblock \href {http://dx.doi.org/10.1093/acprof:oso/9780199677467.001.0001}
  {\path{doi:10.1093/acprof:oso/9780199677467.001.0001}}.

\bibitem{note_bold}
Throughout the manuscript we use the bold face to denote the one-dimensional
  momentum vectors ${\bf k}$ and ${\bf q}$ in order to distinguish them from
  the combined momentum and frequency indices $k=({\bf k},\nu_n)$ and $q=({\bf
  q},\omega_n)$.

\bibitem{doi:10.1021/ja00134a022}
P.~C. Hiberty, D.~Danovich, A.~Shurki, S.~Shaik, Why does benzene possess a d6h
  symmetry? a quasiclassical state approach for probing .pi.-bonding and
  delocalization energies, Journal of the American Chemical Society 117~(29)
  (1995) 7760--7768.
\newblock \href {http://dx.doi.org/10.1021/ja00134a022}
  {\path{doi:10.1021/ja00134a022}}.

\bibitem{Bickers_ModPhysB}
N.~Bickers, Parquet equations for numerical self-consistent-field theory,
  International Journal of Modern Physics B 05~(01n02) (1991) 253--270.
\newblock \href {http://dx.doi.org/10.1142/S021797929100016X}
  {\path{doi:10.1142/S021797929100016X}}.

\bibitem{Diatlov1957}
I.~T. Diatlov, V.~V. Sudakov, K.~A. Ter-Martirosian, Asymptotic meson-meson
  scattering theory, Sov. Phys. JETP 5 (1957) 631.

\bibitem{DeDominicis1964b}
C.~{De Dominicis}, P.~C. Martin,
  \href{http://aip.scitation.org/doi/abs/10.1063/1.1704064}{Stationary entropy
  principle and renormalization in normal and superfluid systems. {II}.
  diagrammatic formulation}, J. Math. Phys. 5~(1) (1964) 31.
\newblock \href {http://dx.doi.org/10.1063/1.1704064}
  {\path{doi:10.1063/1.1704064}}.
\newline\urlprefix\url{http://aip.scitation.org/doi/abs/10.1063/1.1704064}

\bibitem{Wentzell2016}
N.~Wentzell, G.~Li, A.~Tagliavini, C.~Taranto, G.~Rohringer, K.~Held,
  A.~Toschi, S.~Andergassen, High-frequency asymptotics of the vertex function:
  diagrammatic parametrization and algorithmic implementation, ArXiv
  e-prints\href {http://arxiv.org/abs/1610.06520} {\path{arXiv:1610.06520}}.

\bibitem{Held2014}
K.~Held, \href{http://www.cond-mat.de/events/correl14}{Autumn {S}chool on
  {C}orrelated {E}lectrons. {DMFT} at 25: {I}nfinite {D}imensions}, Vol.~4 of
  Modeling and Simulations, Forschungszentrum J{\"u}lich, 2014, Ch. Dynamical
  vertex approximation, [arXiv:1411.5191].
\newline\urlprefix\url{http://www.cond-mat.de/events/correl14}

\bibitem{Rohringer2012}
G.~Rohringer, A.~Valli, A.~Toschi,
  \href{http://link.aps.org/doi/10.1103/PhysRevB.86.125114}{Local electronic
  correlation at the two-particle level}, Phys. Rev. B 86 (2012) 125114.
\newblock \href {http://dx.doi.org/10.1103/PhysRevB.86.125114}
  {\path{doi:10.1103/PhysRevB.86.125114}}.
\newline\urlprefix\url{http://link.aps.org/doi/10.1103/PhysRevB.86.125114}

\bibitem{Toschi2011}
A.~Toschi, G.~Rohringer, A.~Katanin, K.~Held,
  \href{http://dx.doi.org/10.1002/andp.201100036}{Ab initio calculations with
  the dynamical vertex approximation}, Annalen der Physik 523~(8-9) (2011) 698.
\newblock \href {http://dx.doi.org/10.1002/andp.201100036}
  {\path{doi:10.1002/andp.201100036}}.
\newline\urlprefix\url{http://dx.doi.org/10.1002/andp.201100036}

\bibitem{Rohringer2013a}
G.~Rohringer, New routes towards a theoretical treatment of nonlocal electronic
  correlations, Ph.D. thesis, Vienna University of Technology (2013).

\bibitem{Note1}
Evaluated for the $N_{\protect \hspace {-.2em}f}$ frequencies of the inner
  frequency box.

\bibitem{HoerbingerPA}
F.~H\"orbinger, Bachelor Thesis, TU Wien, 2015.

\bibitem{Note2}
Here, we use the interaction expansion continuous time Monte Carlo (CT-QMC)
  method\cite {Rubtsov2005,Gull2011a} in the implementation of Gang Li.

\bibitem{Valli2013}
A.~Valli, Electronic correlations at the nanoscale, Ph.D. thesis, Vienna
  University of Technology (2013).

\bibitem{Muller-Hartmann1988}
E.~M{\"u}ller-Hartmann, \href{https://doi.org/10.1007/BF01311397}{Correlated
  fermions on a lattice in high dimensions}, Zeitschrift f{\"u}r Physik B
  Condensed Matter 74~(4) (1989) 507--512.
\newblock \href {http://dx.doi.org/10.1007/BF01311397}
  {\path{doi:10.1007/BF01311397}}.
\newline\urlprefix\url{https://doi.org/10.1007/BF01311397}

\bibitem{Schaefer15}
T.~Sch\"afer, A.~Toschi, J.~M. Tomczak,
  \href{http://link.aps.org/doi/10.1103/PhysRevB.91.121107}{Separability of
  dynamical and nonlocal correlations in three dimensions}, Phys. Rev. B 91
  (2015) 121107.
\newblock \href {http://dx.doi.org/10.1103/PhysRevB.91.121107}
  {\path{doi:10.1103/PhysRevB.91.121107}}.
\newline\urlprefix\url{http://link.aps.org/doi/10.1103/PhysRevB.91.121107}

\bibitem{PhysRevBselfenergy}
P.~Pudleiner, T.~Sch\"afer, D.~Rost, G.~Li, K.~Held, N.~Bl\"umer, Momentum
  structure of the self-energy and its parametrization for the two-dimensional
  {H}ubbard model, Phys. Rev. B 93 (2016) 195134.
\newblock \href {http://dx.doi.org/10.1103/PhysRevB.93.195134}
  {\path{doi:10.1103/PhysRevB.93.195134}}.

\bibitem{Katanin2013}
A.~A. Katanin, \href{http://stacks.iop.org/1751-8121/46/i=4/a=045002}{The
  effect of six-point one-particle reducible local interactions in the dual
  fermion approach}, J. Phys. A: Math. Theor. 46~(4) (2013) 045002.
\newline\urlprefix\url{http://stacks.iop.org/1751-8121/46/i=4/a=045002}

\bibitem{Ribic2017b}
T.~Ribic, P.~Gunacker, S.~Iskakov, M.~Wallerberger, G.~Rohringer, A.~N.
  Rubtsov, E.~Gull, K.~Held,
  \href{https://link.aps.org/doi/10.1103/PhysRevB.96.235127}{Role of
  three-particle vertex within dual fermion calculations}, Phys. Rev. B 96
  (2017) 235127.
\newblock \href {http://dx.doi.org/10.1103/PhysRevB.96.235127}
  {\path{doi:10.1103/PhysRevB.96.235127}}.
\newline\urlprefix\url{https://link.aps.org/doi/10.1103/PhysRevB.96.235127}

\bibitem{Note3}
For the Pad\'{e}{} interpolation, about $15$ grid points equally spaced with
  respect to Matsubara frequency are utilized and chosen such that the
  high-frequency behavior is not included. The correct tail is anyway
  automatically obtained by the assumed continued fraction.

\bibitem{Godby1988}
R.~W. Godby, M.~Schl\"uter, L.~J. Sham,
  \href{https://link.aps.org/doi/10.1103/PhysRevB.37.10159}{Self-energy
  operators and exchange-correlation potentials in semiconductors}, Phys. Rev.
  B 37 (1988) 10159--10175.
\newblock \href {http://dx.doi.org/10.1103/PhysRevB.37.10159}
  {\path{doi:10.1103/PhysRevB.37.10159}}.
\newline\urlprefix\url{https://link.aps.org/doi/10.1103/PhysRevB.37.10159}

\bibitem{Held2011}
K.~Held, C.~Taranto, G.~Rohringer, A.~Toschi,
  \href{http://arxiv.org/abs/1109.3972}{Hedin equations, {GW}, {GW+DMFT}, and
  all that}Lecture Notes of the Autumn School 2011 Hands-on {LDA+DMFT},
  Forschungszentrum Juelich GmbH (publisher).
\newblock \href {http://arxiv.org/abs/1109.3972} {\path{arXiv:1109.3972}}.
\newline\urlprefix\url{http://arxiv.org/abs/1109.3972}

\bibitem{Note4}
Note that the broadening of the PA spectral function in the top panel \par is
  larger than for each $\protect \mathbf {k}$-resolved component, since the
  Pad\'{e}{} fit was performed after the $\protect \mathbf {k}$-summation.

\bibitem{Note5}
The weight of the spectral features depends slightly on the details of the
  Pad\'{e}{} fit procedure.

\bibitem{Note6}
Please note that applying an electric field within the plane of the benzene
  ring results in a much more complicated equation than Eq.~\ref {eq:ccCkw}
  since it breaks translational invariance along the ring and momentum is not
  conserved any longer.

\bibitem{Note7}
We further checked that the peaks calculated by the Pad\'{e}{} analytical
  continuation reproduce our data when transformed back to imaginary frequency
  where each peak gives a Lorentzian with different weight (weight of the peak
  in real frequencies) and width (energy of the peak).

\bibitem{Koch1972}
E.~Koch, A.~Otto,
  \href{http://www.sciencedirect.com/science/article/pii/0009261472900115}{Optical
  absorption of benzene vapour for photon energies from 6 {eV} to 35 {eV}},
  Chemical Physics Letters 12~(3) (1972) 476 -- 480.
\newblock \href
  {http://dx.doi.org/https://doi.org/10.1016/0009-2614(72)90011-5}
  {\path{doi:https://doi.org/10.1016/0009-2614(72)90011-5}}.
\newline\urlprefix\url{http://www.sciencedirect.com/science/article/pii/0009261472900115}

\bibitem{Lassettre1968}
E.~N. Lassettre, A.~Skerbele, M.~A. Dillon, K.~J. Ross,
  \href{https://doi.org/10.1063/1.1668178}{High-resolution study of
  electron-impact spectra at kinetic energies between 33 and 100 {eV} and
  scattering angles to 16$^\circ$}, J. Chem. Phys. 48~(11) (1968) 5066--5096.
\newblock \href {http://arxiv.org/abs/https://doi.org/10.1063/1.1668178}
  {\path{arXiv:https://doi.org/10.1063/1.1668178}}, \href
  {http://dx.doi.org/10.1063/1.1668178} {\path{doi:10.1063/1.1668178}}.
\newline\urlprefix\url{https://doi.org/10.1063/1.1668178}

\bibitem{Doering2003}
J.~P. Doering, \href{https://doi.org/10.1063/1.1672424}{Low-energy
  electron-impact study of the first, second, and third triplet states of
  benzene}, The Journal of Chemical Physics 51~(7) (1969) 2866--2870.
\newblock \href {http://arxiv.org/abs/https://doi.org/10.1063/1.1672424}
  {\path{arXiv:https://doi.org/10.1063/1.1672424}}, \href
  {http://dx.doi.org/10.1063/1.1672424} {\path{doi:10.1063/1.1672424}}.
\newline\urlprefix\url{https://doi.org/10.1063/1.1672424}

\bibitem{Anisimov1997}
V.~I. Anisimov, A.~I. Poteryaev, M.~A. Korotin, A.~O. Anokhin, G.~Kotliar,
  \href{http://stacks.iop.org/0953-8984/9/7359}{First-principles calculations
  of the electronic structure and spectra of strongly correlated systems:
  dynamical mean-field theory}, Journal of Physics: Condensed Matter 9 (1997)
  7359--7367.
\newline\urlprefix\url{http://stacks.iop.org/0953-8984/9/7359}

\bibitem{Lichtenstein1998}
A.~I. Lichtenstein, M.~I. Katsnelson, Ab initio calculations of quasiparticle
  band structure in correlated systems: {LDA++} approach, Phys. Rev. B 57
  (1998) 6884--6895.
\newblock \href {http://dx.doi.org/10.1103/PhysRevB.57.6884}
  {\path{doi:10.1103/PhysRevB.57.6884}}.

\bibitem{Held2006}
K.~Held, I.~A. Nekrasov, G.~Keller, V.~Eyert, N.~Blümer, A.~K. McMahan, R.~T.
  Scalettar, T.~Pruschke, V.~I. Anisimov, D.~Vollhardt,
  \href{http://dx.doi.org/10.1002/pssb.200642053}{Realistic investigations of
  correlated electron systems with {LDA + DMFT}}, physica status solidi (b)
  243~(11) (2006) 2599--2631, previously appeared as Psi-k Newsletter No. 56
  (April 2003).
\newblock \href {http://dx.doi.org/10.1002/pssb.200642053}
  {\path{doi:10.1002/pssb.200642053}}.
\newline\urlprefix\url{http://dx.doi.org/10.1002/pssb.200642053}

\bibitem{Kotliar2006}
G.~Kotliar, S.~Y. Savrasov, K.~Haule, V.~S. Oudovenko, O.~Parcollet, C.~A.
  Marianetti,
  \href{http://link.aps.org/doi/10.1103/RevModPhys.78.865}{Electronic structure
  calculations with dynamical mean-field theory}, Rev. Mod. Phys. 78 (2006)
  865.
\newblock \href {http://dx.doi.org/10.1103/RevModPhys.78.865}
  {\path{doi:10.1103/RevModPhys.78.865}}.
\newline\urlprefix\url{http://link.aps.org/doi/10.1103/RevModPhys.78.865}

\bibitem{Held2007}
K.~Held, Electronic structure calculations using dynamical mean field theory,
  Advances in Physics 56 (2007) 829--926.
\newblock \href {http://dx.doi.org/10.1080/00018730701619647}
  {\path{doi:10.1080/00018730701619647}}.

\bibitem{Biermann2003}
S.~Biermann, F.~Aryasetiawan, A.~Georges,
  \href{http://link.aps.org/doi/10.1103/PhysRevLett.90.086402}{First-principles
  approach to the electronic structure of strongly correlated systems:
  Combining the {$GW$} approximation and dynamical mean-field theory}, Phys.
  Rev. Lett. 90 (2003) 086402.
\newblock \href {http://dx.doi.org/10.1103/PhysRevLett.90.086402}
  {\path{doi:10.1103/PhysRevLett.90.086402}}.
\newline\urlprefix\url{http://link.aps.org/doi/10.1103/PhysRevLett.90.086402}

\bibitem{Tomczak2012}
J.~M. Tomczak, M.~Casula, T.~Miyake, F.~Aryasetiawan, S.~Biermann,
  \href{http://stacks.iop.org/0295-5075/100/i=6/a=67001}{Combined {GW} and
  dynamical mean-field theory: Dynamical screening effects in transition metal
  oxides}, EPL (Europhysics Letters) 100~(6) (2012) 67001.
\newline\urlprefix\url{http://stacks.iop.org/0295-5075/100/i=6/a=67001}

\bibitem{jmt_sces14}
J.~M. Tomczak,
  \href{http://iopscience.iop.org/article/10.1088/1742-6596/592/1/012055}{{QS{\it
  GW}+DMFT}: an electronic structure scheme for the iron pnictides and beyond},
  J. Phys.: Conference Series 592~(1) (2015) 012055.
\newline\urlprefix\url{http://iopscience.iop.org/article/10.1088/1742-6596/592/1/012055}

\bibitem{Galler2017b}
A.~Galler, P.~{Thunstr{\"o}m}, J.~{Kaufmann}, M.~{Pickem}, J.~M. {Tomczak},
  K.~{Held}, \href{https://arxiv.org/abs/1710.06651}{{The AbinitioD$\Gamma$A
  Project v1.0: Non-local correlations beyond and susceptibilities within
  dynamical mean-field theory}}, arXiv:1710.06651\href
  {http://arxiv.org/abs/1710.06651} {\path{arXiv:1710.06651}}.
\newline\urlprefix\url{https://arxiv.org/abs/1710.06651}

\bibitem{Rubtsov2005}
A.~N. Rubtsov, V.~V. Savkin, A.~I. Lichtenstein,
  \href{http://link.aps.org/doi/10.1103/PhysRevB.72.035122}{Continuous-time
  quantum {Monte} {Carlo} method for fermions}, Phys. Rev. B 72 (2005) 035122.
\newblock \href {http://dx.doi.org/10.1103/PhysRevB.72.035122}
  {\path{doi:10.1103/PhysRevB.72.035122}}.
\newline\urlprefix\url{http://link.aps.org/doi/10.1103/PhysRevB.72.035122}

\end{thebibliography}

\end{document}